\documentclass[11pt,a4paper]{article}
\pdfoutput=1
\usepackage{jheppub}

\usepackage{color}
\usepackage{amsmath}
\usepackage{pifont}   
\usepackage{verbatim}       
\usepackage{acronym} 
    
\usepackage{amsfonts}
\usepackage{amssymb}
\usepackage{mathrsfs} 
\usepackage{graphicx}
\usepackage{multirow} 
\usepackage{slashed} 
\usepackage{array}

\usepackage{graphicx}
\usepackage{epsfig}
\usepackage{lscape}
\usepackage{rotating}
\usepackage{epsf}
\usepackage{bm}
\usepackage{booktabs}
\usepackage{array}
\usepackage{caption}
\usepackage{subcaption}
\usepackage[utf8]{inputenc}
\usepackage{float}

\usepackage{multirow}
\usepackage{array,booktabs,colortbl,multirow}
\usepackage{colortbl,xcolor,colordvi,color}
\usepackage{rotating}
\allowdisplaybreaks

\newcommand{\beq}{\begin{align}}
\newcommand{\eeq}{\end{align}}
\def\be{\begin{equation}}
\def\ee{\end{equation}}
\def\bea{\begin{eqnarray}}
\def\eea{\end{eqnarray}}
\def\bitem{\begin{itemize}}
\def\eitem{\end{itemize}}
\newcommand{\bec}{\begin{center}}
\newcommand{\eec}{\end{center}}
\newcommand{\ba}{\begin{array}}
\newcommand{\ea}{\end{array}}

\title{Dark matter in Hidden Valley models with stable and unstable light dark mesons}

\author[1, 2]{Hugues Beauchesne,}
\author[1]{Enrico Bertuzzo}
\author[1]{and Giovanni Grilli di Cortona}

\affiliation[1]{Instituto de F\'isica, Universidade de S\~ao Paulo, \\C.P. 66.318, 05315-970 S\~ao Paulo, Brazil}
\affiliation[2]{Department of Physics, Ben-Gurion University, \\Beer-Sheva 8410501, Israel}

\emailAdd{beauches@post.bgu.ac.il, bertuzzo@if.usp.br, ggrilli@if.usp.br}

\abstract{It is a distinct possibility that a Hidden Valley sector would have a spectrum of light particles consisting of both stable and unstable dark mesons. The simultaneous presence of these two types of particles can lead to novel mechanisms for generating the correct dark matter relic abundance, which in turn can reflect themselves into new exotic signatures at colliders. We study the viability of such sectors for various Hidden Valley models and map the valid parameter space to possible collider signatures. Mediators studied include various scalar bifundamentals and a heavy $Z'$. It is shown that in general bounds from direct and indirect detection can easily be avoided. In most of the allowed parameter space, the relic density is determined by stable mesons annihilating to unstable ones which in turn decay quickly to Standard Model particles. Dark mesons that decay mainly to heavier Standard Model fermions allow for more valid parameter space, though dark mesons are still allowed to decay exclusively to some of the lighter fermions. Possible exotic collider signatures include displaced vertices, emerging jets and semivisible jets.}

\begin{document}

\maketitle
\section{Introduction}\label{Sec:Intro}
The existence of dark matter (DM) is for all intents and purposes an established fact. A multitude of astrophysical observations such as anisotropies in the Cosmic Microwave Background (CMB), large scale structures of the Universe and galactic rotational curves all point toward the presence of a long-lived and non-baryonic dark matter component with a density roughly five times that of visible matter \cite{Aghanim:2018eyx}. For the past few decades, the leading dark matter candidates have been Weakly Interacting Massive Particles (WIMP), which are assumed to generate the correct relic abundance via thermal freeze-out \cite{Lee:1977ua, Vysotsky:1977pe, Kolb:1990vq}.

In contrast to astrophysical experiments, direct detection experiments have failed to convincingly detect any dark matter particles, and in fact have imposed stringent constraints on their interactions with baryonic matter \cite{Aprile:2017iyp}. This has put increasing tension on the WIMP paradigm and lead many to consider alternative dark matter scenarios. These include Strongly Interacting Massive Particles (SIMP) \cite{Hochberg:2014dra}, Elastically Decoupling Dark Matter (ELDER) \cite{Kuflik:2015isi} and codecaying dark matter \cite{Dror:2016rxc, Farina:2016llk, Okawa:2016wrr}, to name just a few.

One interesting and understudied possibility is that dark matter consists of the stable dark mesons of a Hidden Valley sector \cite{Strassler:2006im} that also contains unstable ones. Such a dark sector is certainly not a far-fetched idea. For example, had the electrons and the muons been heavier than the charged pion, the latter would have been stable while the neutral pion would still be able to decay to two photons. Had the Cabbibo-Kobayashi-Maskawa matrix been diagonal, some of the kaons would also have been stable. Such sectors can generate the correct dark matter relic density via novel mechanisms and as such can potentially avoid the bounds from direct detection, while also leading to novel cosmological and collider signatures. See Ref.~\cite{Okawa:2016wrr, Hochberg:2018vdo, Alves:2010dd,Lee:2015gsa} for previous work on these sectors.\footnote{Stable dark baryons can potentially contribute to the dark matter relic density. However, their relic density turns out to be negligibly small in dark QCD sectors without asymmetry \cite{Chivukula:1989qb}. Dark baryon asymmetric dark matter, though certainly a valid option, is beyond the scope of this article \cite{Petraki:2013wwa, Kaplan:2009ag, Zurek:2013wia}.}

With this context in mind, we present an overview of the cosmological constraints on Hidden Valley models whose spectra of light particles consists of both stable and unstable dark mesons. More precisely, we apply bounds from direct and indirect detection to a set of Hidden Valley benchmark models. For dark mesons below a few hundred GeV, most of the valid parameter space falls into a regime which we refer to as \emph{coupling-independent}. In this regime, interactions between the two sectors are weak enough that annihilation of dark pions to Standard Model (SM) particles decouples at very early times. The process that is then responsible for reducing the dark matter density is the annihilation of stable dark pions to unstable ones that subsequently decay to the Standard Model. However, the interactions between the two sectors are still sufficiently strong in this regime for unstable dark pions to maintain their thermal equilibrium density via decay and inverse decay until after the dark pions have decoupled from each other. What determines the dark matter relic density is then mainly when the dark pions decouple from each other, which is controlled by the pion decay constant. The dark matter relic density is then mainly independent of the interactions between the Standard Model and dark sector, hence the name coupling-independent. This allows the bounds from direct detection to be naturally avoided and massively expands the range of valid parameter space.

Additionally, we map the allowed parameter space to collider signatures, which can include exotic signals such as displaced vertices, emerging jets \cite{Schwaller:2015gea} and semivisible jets \cite{Cohen:2015toa}. We also show that indirect detection searches could soon probe large regions of parameter space. Because this is where these exotic signatures are present and where the mechanism functions most naturally, we focus on dark mesons ranging from a few GeV to a few hundred GeV. In many ways, the present article is a continuation of Ref.~\cite{Beauchesne:2017yhh} and shows how the models presented in that article can explain the dark matter relic abundance.

This article is organized as follows. First, we present a more precise definition of our benchmark confining sector and discuss more carefully the mechanism to generate the correct dark matter relic density. Then, we provide an overview of the different experimental constraints applied in the analysis. With this done, we present an overview of the experimental bounds for different Hidden Valley models with scalar or vector mediators ($Z'$). This is accompanied by a list of the associated collider signatures. Additional discussions are then presented. Finally, the appendix includes a description of the procedure used to calculate the relic densities.

\section{General setup and overview of the mechanism}\label{Sec:Setup&Overview}
We begin by presenting the benchmark dark sector that will be used throughout the article and explain the mechanism that will lead to the correct dark matter relic abundance. We refer to appendix~\ref{ap:Technicalities} and the articles cited in the present section for the technical details of the different mechanisms involved.

In practice, it is inconceivable to constrain every possible dark sector and a benchmark must be chosen. We therefore decide to concentrate on a dark sector that is a copy of QCD with three light dark quarks. This choice is justified as it is a familiar example, is present in several theories (e.g. Mirror Twin Higgs \cite{Chacko:2005pe, Barbieri:2005ri}) and allows for the presence of the Wess-Zumino-Witten (WZW) term \cite{Wess:1971yu, Witten:1983tw, Witten:1983tx}. This term plays a crucial role in certain regions of parameter space, but its effect can be made to vanish by raising sufficiently the mass of a single dark quark.

More precisely, we assume the presence of a new confining group $\mathcal{G}$, which we take to be $SU(3)$. We assume the dark sector to consist of a set of three Dirac fermion dark quarks $n_i$, where $i$ runs from 1 to 3. They are assumed fundamentals of $\mathcal{G}$. As in the Standard Model, there is an approximate $SU(3)_L\times SU(3)_R$ symmetry that is assumed to be broken spontaneously to the diagonal $SU(3)_V$ by the dark quark condensate. We will consider the possibility of the latter symmetry being broken explicitly by the dark quark masses. The spontaneously broken axial symmetry results in a set of pseudo-Goldstone bosons, which we refer to in general as dark pions. We refer to the stable ones as $\pi^s_i$ and the unstable ones as $\pi^u_i$, where $i$ serves as a label. In all cases studied below, there will be two unstable real dark pions and three stable complex dark pions. The pion matrix can be written as:\footnote{Different quark masses can lead to mixing between $\pi^u_1$ and $\pi^u_2$. We take this effect into account in our code.}
\begin{equation}\label{eq:PionMatrix}
  \Pi=
  \begin{pmatrix}
    \frac{1}{\sqrt{2}}\pi^u_1 + \frac{1}{\sqrt{6}}\pi^u_2 & \pi^s_1 & \pi^s_2 \\
    \bar{\pi}^s_1 & -\frac{1}{\sqrt{2}}\pi^u_1 + \frac{1}{\sqrt{6}}\pi^u_2 & \pi^s_3 \\
    \bar{\pi}^s_2 & \bar{\pi}^s_3 & -\sqrt{\frac{2}{3}}\pi^u_2
  \end{pmatrix}.
\end{equation}
A set of three $U(1)_i$ symmetries can be defined under which only an individual $n_i$ is charged at a time. The $\pi^s_i$'s are then non-trivially charged under a subset of these $U(1)_i$ symmetries. These symmetries are sufficient to insure the stability of at least two $\pi^s_i$'s. They are however insufficient to insure the stability of all three $\pi^s_i$'s, as the $U(1)_i$ charges of a given $\pi^s_i$ can be expressed as a linear combination of the charges of two other $\pi^s_i$'s. All three $\pi^s_i$'s will however be stable if the mass splitting between them is insufficient to allow the decay of a $\pi^s_i$ to two others. We will assume this to be the case from now on. In contrast, the $\pi^u_i$'s are neutral under the different $U(1)_i$'s and their stability is therefore not insured by these symmetries. The $SU(3)_V$ left unbroken by the quark condensate could potentially leave the $\pi^u_i$'s stable, but this is by definition outside the scope of our article and the models we will introduce in Sec.~\ref{Sec:Models} will explicitly break this symmetry. 

\begin{figure}[t!]
  \centering
  \begin{subfigure}{0.48\textwidth}
    \centering
    \includegraphics[width=\textwidth, viewport = 0 0 450 264]{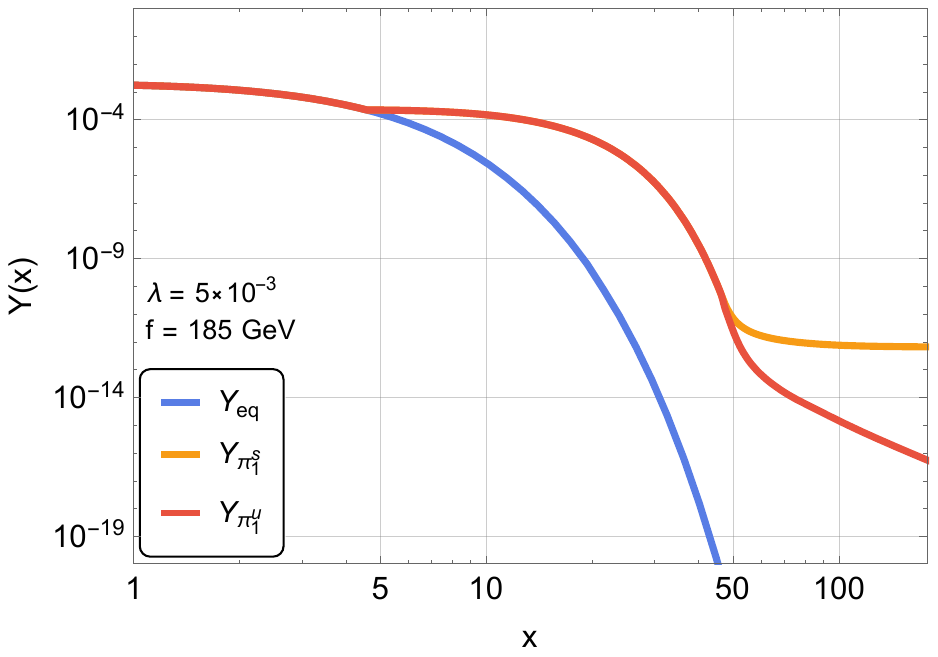}
    \caption{}
    \label{fig:EvolutionRelicDensitya}
  \end{subfigure}
  ~
    \begin{subfigure}{0.48\textwidth}
    \centering
    \includegraphics[width=\textwidth, viewport = 0 0 450 264]{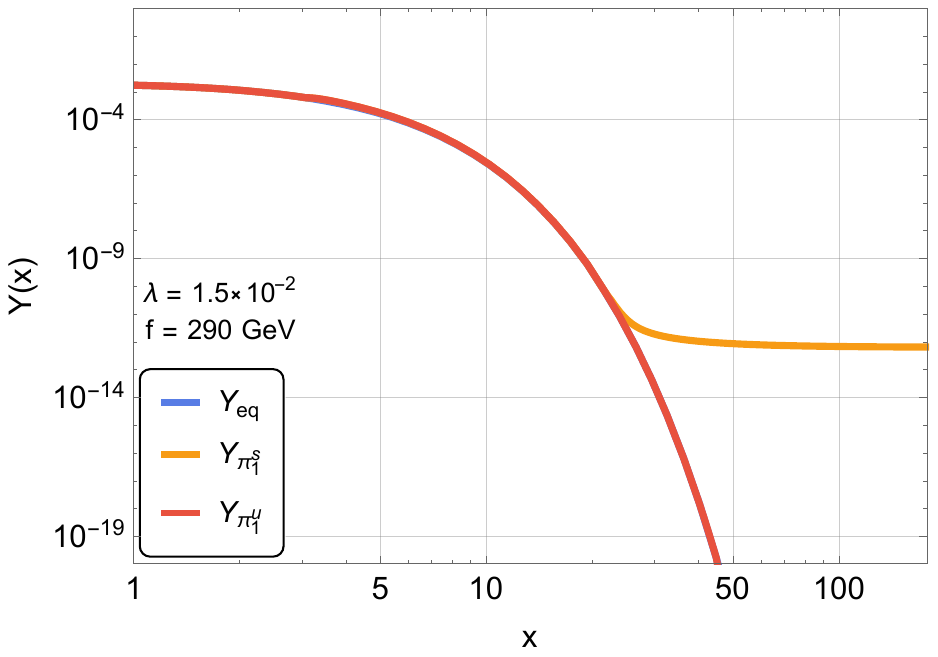}
    \caption{}
    \label{fig:EvolutionRelicDensityb}
  \end{subfigure}
\caption{Evolution of the number density per entropy density of dark pions for the model of Sec.~\ref{sSec:ModelII} and different parameters. Dark pion masses are all set to 100 GeV and the mediator is set to 1 TeV. The dark quark masses $m_{n_i}$ and $B_0$ are all set equal. The couplings are taken as $\lambda^S_{D^c_{ijk}} = \lambda\delta_{i1}\delta_{jk}$. (a)~The parameter $\lambda$ is set to $5\times 10^{-3}$ and $f$ is adjusted to 185~GeV to reproduce the correct DM abundance. (b)~The parameter $\lambda$ is set to $1.5\times 10^{-2}$ and $f$ is adjusted to 290~GeV to reproduce the correct DM abundance. See Sec.~\ref{sSec:ModelII} for more details.}\label{fig:EvolutionRelicDensity}
\end{figure}

In principle, mesonic dark matter can reproduce the correct dark matter relic density via several mechanisms. Assume an effective operator that controls the interactions between the dark sector and the Standard Model with coefficient $\lambda$. For $\lambda$ going to zero, decay or annihilation of dark pions to the Standard Model are strongly suppressed and the two sectors decouple at small $x\equiv m_{\pi^s_1}/T$, where $m_{\pi^s_1}$ is the mass of $\pi^s_1$ and $T$ the temperature of the Standard Model sector. Number-changing processes involving only dark pions are therefore the dominant factor in determining the relic density. These are mainly $3\to 2$ processes, which are a consequence of the WZW term. This regime corresponds either to SIMP, if the two sectors maintain kinetic equilibrium until $3\to 2$ processes have frozen-out, or ELDER otherwise. The dark matter relic density is then mostly determined by when the $3\to 2$ processes become inefficient, albeit the evolution of the temperature of the dark sector can have non-negligible effects. Because of the already extensive literature on the subject, we will not study this limit.

As $\lambda$ increases, the system eventually enters the codecaying dark matter regime. The dark pions continue to decouple from the Standard Model sector at small $x$. However, because of a suppression by a larger power of the number density, $3\to 2$ processes are less important for the determination of the final dark matter relic abundance than the annihilation of two stable pions to two unstable ones. These subsequently decay to the Standard Model and are not replaced as inverse decay is not efficient in this regime. The overall dark pion density continues to decrease until they decouple from each other. The relic density is then dependent on both $\lambda$ and the strength with which pions interact between themselves. The evolution of the relic density is shown in Fig.~\ref{fig:EvolutionRelicDensitya} for the model which will be presented in Sec.~\ref{sSec:ModelII}.

As $\lambda$ continues to increase, it will eventually reach a point where the number densities of the $\pi^u_i$'s retain their equilibrium values until after the dark pions have decoupled from each other. This is shown in Fig.~\ref{fig:EvolutionRelicDensityb}. At this point, the exact value of $\lambda$ only has a marginal effect on the final dark matter relic density. In fact, $\lambda$ can be changed by orders of magnitudes in some cases without having any significant impact on the dark matter relic density. Though never thoroughly studied, this regime was mentioned in Ref.~\cite{Okawa:2016wrr}. It will turn out that most of the unexcluded parameter space corresponds to this regime for dark pions below a TeV. For convenience, we will refer to this regime as the `coupling-independent' regime. 

Finally, for very large $\lambda$, annihilation of two stable dark pions to SM particles becomes important. This is simply the usual thermal freeze-out process of WIMPs. In many cases presented below, this region will be almost completely excluded by direct detection.

\begin{figure}[t]
\begin{center}
    \includegraphics[width=0.6\textwidth, viewport = 50 0 450 298]{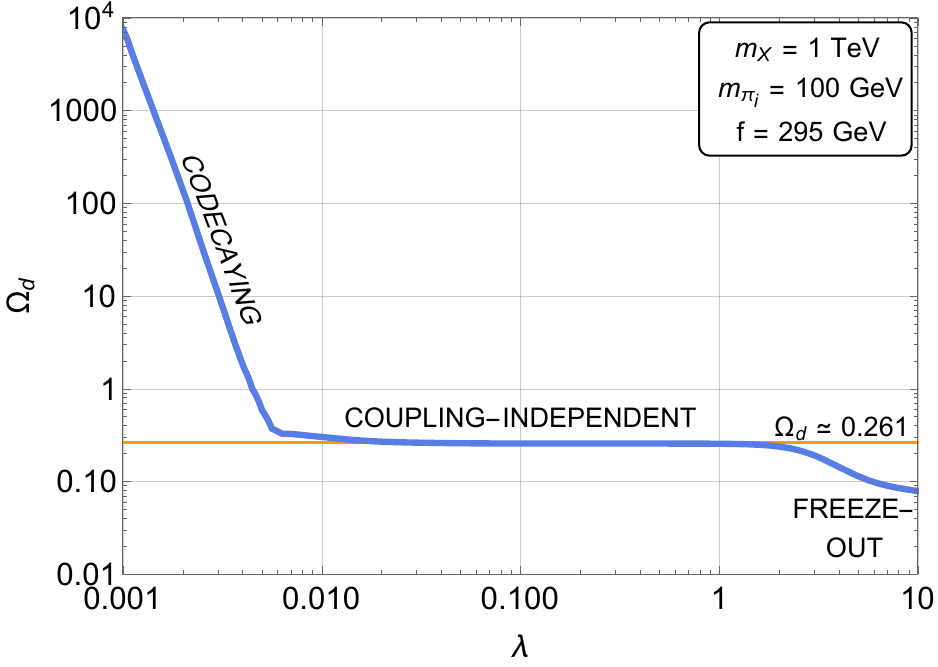}
\end{center} 
\caption{Dark matter relic abundance for the model of Sec.~\ref{sSec:ModelII} as a function of $\lambda$. Dark pion masses are all set to 100 GeV and the mediator is set to 1 TeV. The dark quark masses $m_{n_i}$ and $B_0$ are all set equal. The couplings are taken as $\lambda^S_{D^c_{ijk}} = \lambda\delta_{i1}\delta_{jk}$. The pion decay constant is set to 295~GeV so that the flat section corresponds approximatively to the correct DM abundance. See Sec.~\ref{sSec:ModelII} for more details.}\label{fig:OmegavslambdaR}
\end{figure}

Fig.~\ref{fig:OmegavslambdaR} summarizes the evolution of the dark matter relic density as a function of $\lambda$. The model of Sec.~\ref{sSec:ModelII} was used. The three possible regimes can be clearly seen. They are from left to right: codecaying dark matter, coupling-independent and standard thermal freeze-out.

\section{Overview of constraints}\label{Sec:Constraints}
We present in this section the procedures through which constraints are applied. In this analysis, we used the experimental value for the relic abundance $\Omega_d h^2=0.1200\pm0.0012$ at $68\%$ confidence \cite{Aghanim:2018eyx}. The results for the different benchmark models will appear in Sec.~\ref{Sec:Models}.

\subsection{Direct detection}\label{sSec:DirectDetection}
In scenarios where the stable dark pions $\pi_i^s$ couple with the SM quarks, we can have dark pions recoil on nucleons. As a consequence, direct detection experiments can potentially set an upper bound on the coupling of the stable dark pions with quarks. 

In general, direct detection rates depend on the dark matter local energy density $\rho_{\mathrm{loc}} = \sum_i^n \rho_i$, where $\rho_{\mathrm{loc}} \sim 0.3$ GeV/cm$^3$ and $\rho_i$ are the individual local energy densities for the different stable dark pions $\pi_i^s$. Properly taking into account direct detection constraints in multicomponent dark matter can in principle be a non-trivial task. This is because even if heavier DM particles have a smaller number density, they probe larger velocity distributions. However, we will only be concerned with small mass splittings, in which case this subtlety does not apply.\footnote{We only consider small splittings for two reasons. First, symmetry will force some of the dark pions to be almost degenerate in mass, in a similar way to isospin in the Standard Model. Second, a dark pion that is considerably heavier than the others is generally irrelevant to dark matter considerations because of Boltzmann suppression.} Furthermore, we will assume the proportionality of the global and local densities, or so-called proportionality ansatz \cite{Bertone:2010rv}, which is a standard assumption of multicomponent dark matter (see for example Refs.~\cite{Bertone:2010rv, Bertone:2012fua, Anderhalden:2012qt, Bhattacharya:2013hva, Blennow:2015gta, Herrero-Garcia:2017vrl}). It has been shown to be an excellent approximation for warm dark matter heavier than a few keV \cite{Anderhalden:2012qt}, which is far below the masses we will consider, even for considerable splitting. From a more intuitive point of view, the proportionality hypothesis is expected to hold even more strongly than in previous studies, as the stable dark pions are close in mass, all interact with each other with similar strength and all interact very little with the Standard Model.

In Sec.~\ref{Sec:Models} we will consider constraints on spin independent scattering from the Xenon1T experiment \cite{Aprile:2017iyp}. In computing the direct detection rates, we take into account also the running and mixing of the various operators in going from the scale where the operators are defined (the scale that correspond roughly to the mass of the mediator) to the direct detection scale \cite{Crivellin:2014qxa, DEramo:2014nmf, DEramo:2016gos, Bishara:2017pfq}. Operators that couple directly with the first quark generation will give strong constraints and the running will affect the result sub-dominantly. On the other hand, operators that contain second and third generation quarks will receive their larger contribution from the running and mixing. In our analysis, we numerically implemented the running and mixing of the Standard Model current via the runDM code \cite{DEramo:2016gos,runDM}. Direct detection constraints depend only on the dark matter mass, on the mediator mass and on the coupling of the operator that links the Standard Model and the dark sectors. As a consequence, we expect these bounds to be negligible at small couplings.

All the direct detection bounds correspond to 90\% confidence.

\subsection{Indirect detection}\label{sSec:IndirectDetection}
As $\pi^s_i$'s collide with each others, they produce $\pi^u_i$'s. These in turn decay back to the Standard Model and produce gamma rays or other light particles that can be detected or affect their environment. As such, indirect detection and disturbances in the CMB provide an upper bound on the strength with which $\pi^s_i$'s interact with each others. 

The pair production of $\pi^u_i$'s from collisions of $\pi^s_i$'s leads to cascade decays, albeit with usually only one intermediary state. These cascade decays have been thoroughly treated in Ref.~\cite{Elor:2015bho}, from which we extract our bounds. Constraints from the CMB from Plank \cite{Ade:2015xua}, indirect detection in dwarf galaxies from Fermi-LAT \cite{Drlica-Wagner:2015xua} and positron excess from AMS-02 \cite{Aguilar:2014mma, Accardo:2014lma} are taken into account. The bounds correspond to 95$\%$ confidence. Note that annihilation of two dark pions directly to SM particles can also be constrained by indirect detection, but this is typically overshadowed by bounds from direct detection. Also note that indirect detection bounds overshadow those from dark matter self-interaction \cite{Clowe:2003tk, Markevitch:2003at, Randall:2007ph}.

The indirect detection limits are likely to change in the near future. In the absence of a signal, additional data from Planck, Fermi-LAT and AMS-02 will strengthen the limits but not change their qualitative behaviour \cite{Elor:2015bho}. The possible discovery of new dwarf spheroidal galaxies could lead to an improvement in the limits of about an order of magnitude for Fermi-LAT \cite{Charles:2016pgz}. Stronger limits may be obtained with future more powerful instruments, such as GAMMA-400 \cite{Galper:2013sfa,Cumani:2015ava} or HERD (High Energy cosmic Radiation Detection) \cite{Zhang:2014qga,Huang:2015fca}. Finally, there are good prospects for improvements at high mass (above $\sim100$ GeV) with CTA (Cherenkov Telescope Array) \cite{Consortium:2010bc, Doro:2012xx, Silverwood:2014yza}. 

If the dark pions were degenerate in mass, dark pion scattering would be $p$-wave suppressed. This would result in both much lower thermally averaged cross sections and much less stringent limits on these (see for example Ref.~\cite{Zhao:2016xie}). The end result is that bounds from indirect detection would be much weaker. This degeneracy is however unnatural, as it implies a symmetry that is explicitly broken by the fact that some dark pions are stable and some unstable. We will therefore not consider degenerate dark pions. Typically, one expects the lightest dark pion to be unstable, either because of different dark quark masses or radiative corrections. In this case, the annihilation of two stable pions to the lightest unstable pions is $s$-wave and stronger bounds can be applied from indirect detection. 

In addition, it is important that the decay of the $\pi^u_i$'s do not disturb Big Bang Nucleosynthesis (BBN). Such an issue is avoided if the $\pi^u_i$'s have decayed by the time BBN starts. As such, we will require that the lifetime of the $\pi^u_i$'s be inferior to 0.1 s. This bound is however overshadowed by the bounds on indirect detection.

\section{Constraints for benchmark models}\label{Sec:Models}
In this section, we present different benchmark models for how the dark quarks interact with the Standard Model. The constraints from Sec.~\ref{Sec:Constraints} are applied for each of them and the unconstrained parameter space is mapped to possible collider signatures.

\subsection{Scalar mediators interacting only with down-type quarks}\label{sSec:ModelI}
The first type of models we consider is the benchmark dark sector communicating with the Standard Model via a scalar that couples to both dark quarks and down-type quarks. More precisely, we introduce the operator:
\begin{equation}\label{eq:OpModelI}
  \lambda^S_{D^c_{ij}} (X^S_{D^c})^\dagger \bar{n}_i P_R D^c_j +\text{h.c.}, 
\end{equation}
where $\lambda^S_{D^c_{ij}}$ is a coefficient, $X^S_{D^c}$ the mediator and $D^c_i$ the right-handed down-type quarks. This mediator is a special case of the operators of Category I of Ref.~\cite{Beauchesne:2017yhh}, from which we borrowed and expanded the notation. It has the same SM gauge numbers as $D^c_i$ and is an anti-fundamental of $\mathcal{G}$. We assume a single $\lambda^S_{D^c_{ij}}$ to be non-zero to avoid flavor issues and refer to Ref.~\cite{Renner:2018fhh} for a more complete discussion on flavor constraints. The dark quark that couples with $X^S_{D^c}$ is labelled as $n_1$ and the others as $n_2$ and $n_3$. This choice of $\lambda^S_{D^c_{ij}}$ also insures the stability of all $\pi^s_i$'s, while making the $\pi^u_i$'s unstable. From a model building point of view, having $X^S_{D^c}$ communicate with a single $n_i$ is a point of enhanced symmetry, as this maintains an $SU(2)$ subgroup intact, and is therefore technically natural. In a more complete model like Twin Higgs, couplings with the two other $n_i$'s could be forbidden because of mirror weak hypercharge for example. It is technically natural to have the non-zero $\lambda^S_{D^c_{ij}}$ have any value, as it breaks a $U(1)$ symmetry under which only $X^S_{D^c}$ is charged, in addition to an $SU(3)$ that might be unbroken by the masses of the dark quarks.

\begin{figure}[t!]
  \centering
  \begin{subfigure}{0.48\textwidth}
    \centering
    \includegraphics[width=\textwidth, viewport = 0 0 450 450]{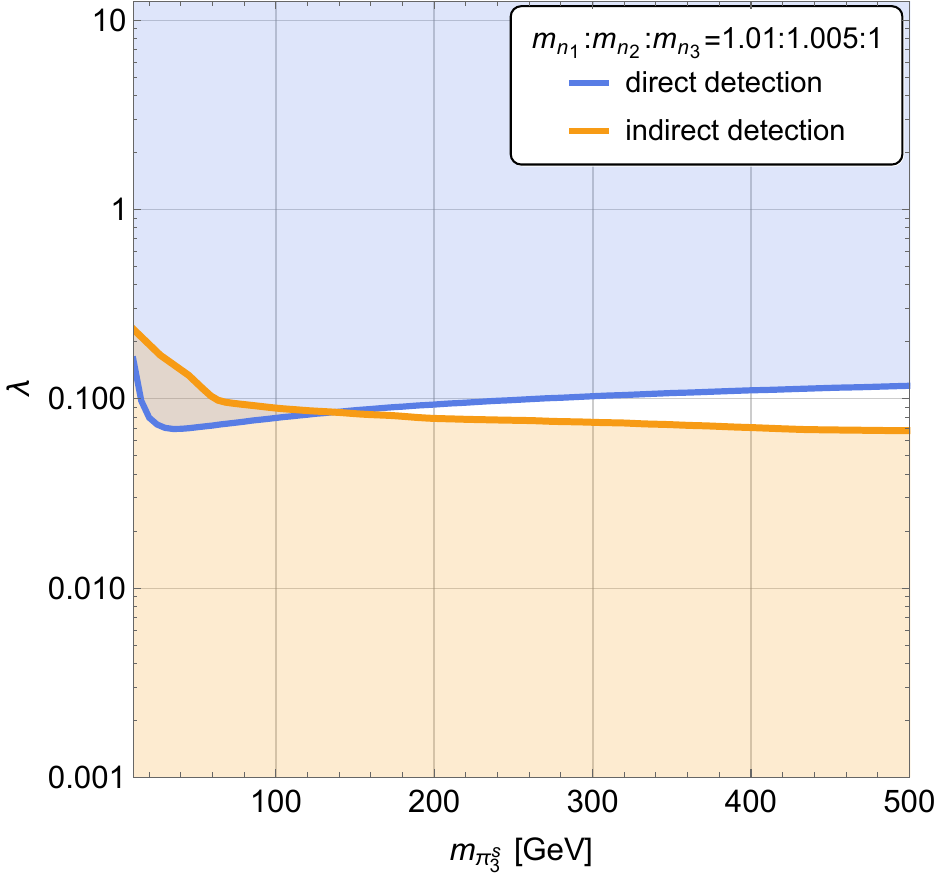}
    \caption{}
     \label{sfig:downa} 
  \end{subfigure}
    ~
    \begin{subfigure}{0.48\textwidth}
    \centering
    \includegraphics[width=\textwidth, viewport = 0 0 450 450]{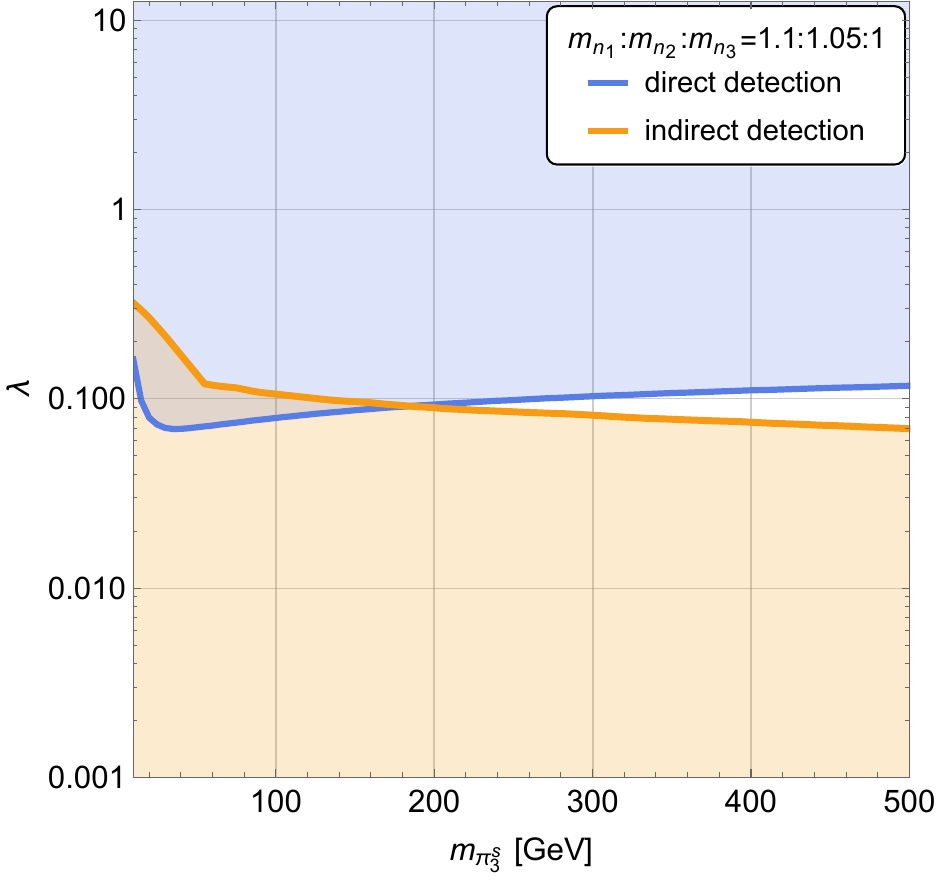}
    \caption{}
     \label{sfig:downb} 
  \end{subfigure}
\caption{Constraints on the coupling $\lambda^S_{D_1^c}$ and the mass of the lightest stable dark pion $m_{\pi_3^s}$. In each point the relic abundance matches the experimental value and $m_{X}=1$ TeV. The blue shaded area is excluded by direct detection experiments, while the orange region is excluded by indirect dark matter searches. (a)~The ratio of the dark quark masses is set to $n_1$:$n_2$:$n_3$=1.01:1.005:1. (b)~The ratio of the dark quark masses is set to $n_1$:$n_2$:$n_3$=1.1:1.05:1.}\label{fig:down}
\end{figure}

\begin{figure}[t!]
  \centering
  \begin{subfigure}{0.48\textwidth}
    \centering
    \includegraphics[width=\textwidth, viewport = 0 0 450 425]{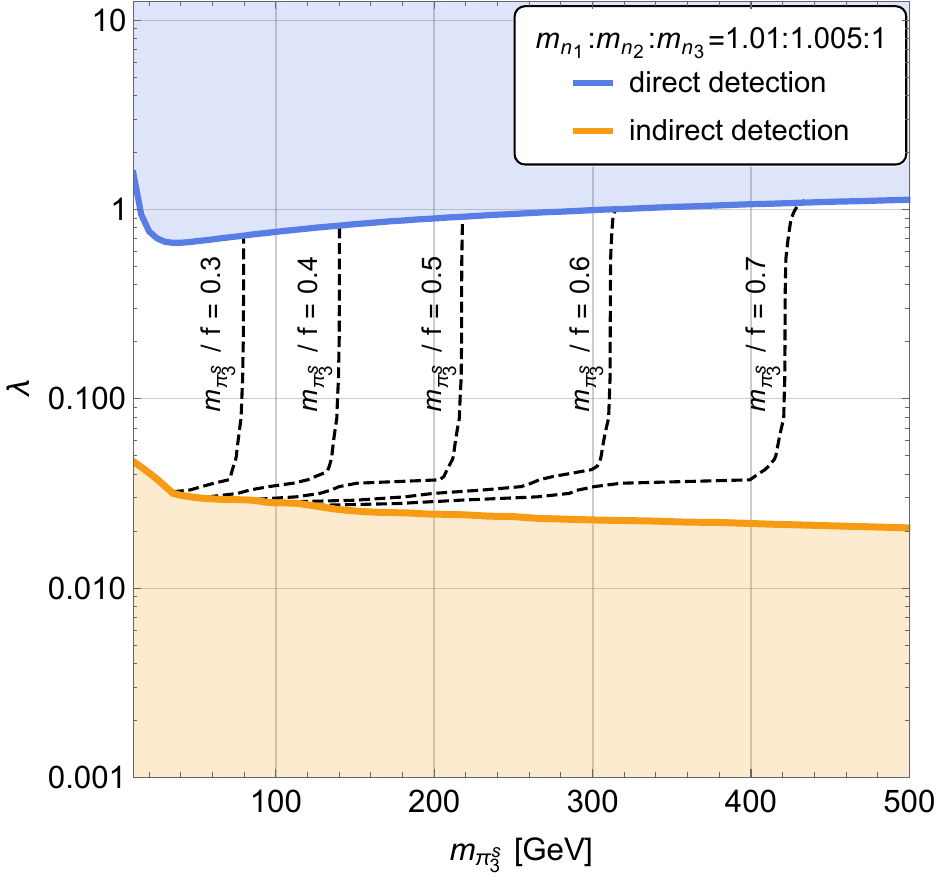}
    \caption{}
    \label{sfig:strangea}
  \end{subfigure}
  ~
    \begin{subfigure}{0.48\textwidth}
    \centering
    \includegraphics[width=\textwidth, viewport = 0 0 450 425]{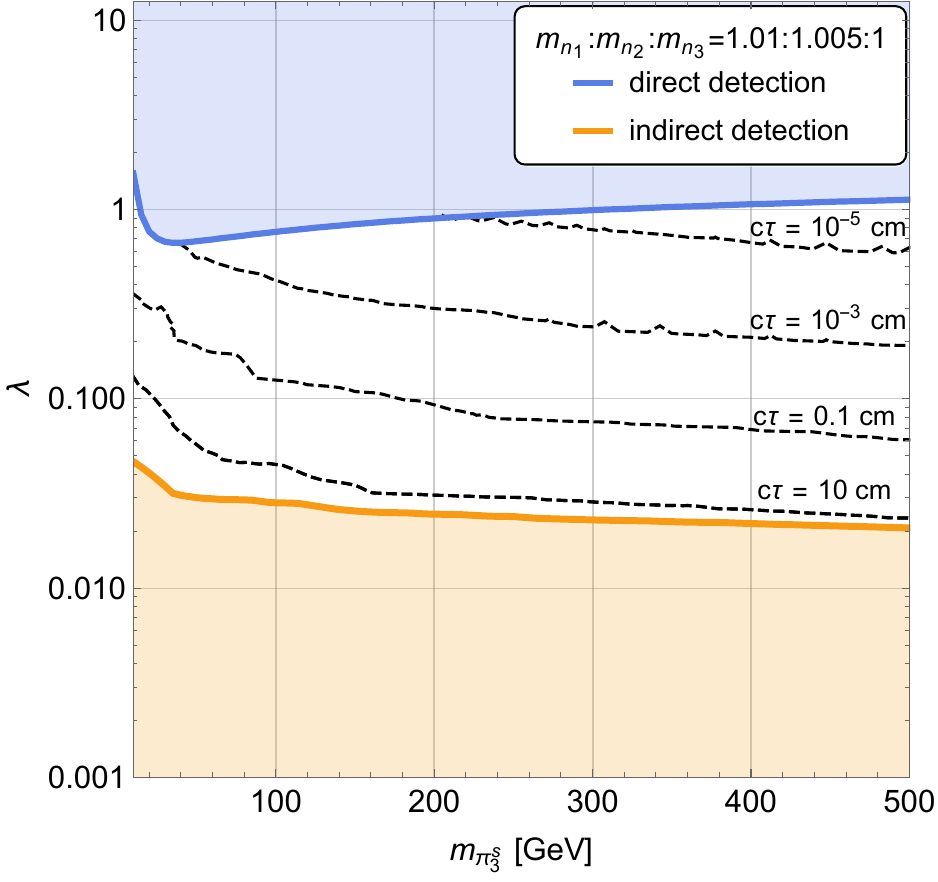}
    \caption{}
    \label{sfig:strangeb}
  \end{subfigure}
  ~
  \begin{subfigure}{0.48\textwidth}
    \centering
    \includegraphics[width=\textwidth, viewport = 0 0 450 425]{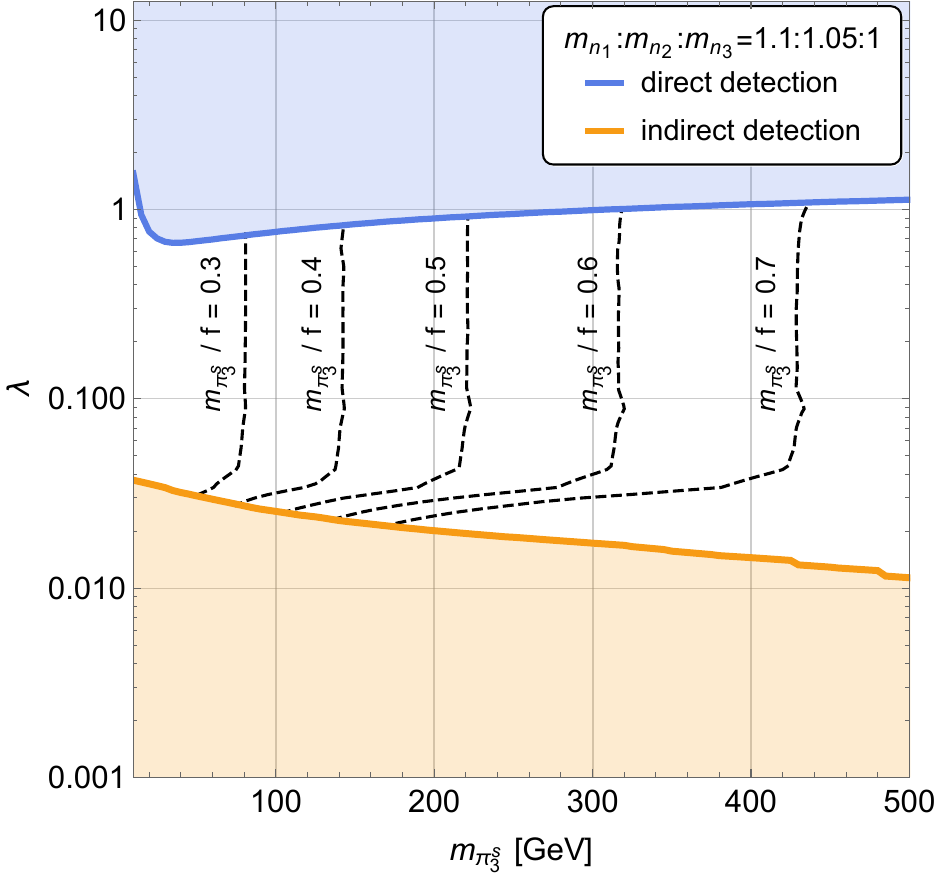}
    \caption{}
    \label{sfig:strangec}
  \end{subfigure}
  ~
    \begin{subfigure}{0.48\textwidth}
    \centering
    \includegraphics[width=\textwidth, viewport = 0 0 450 425]{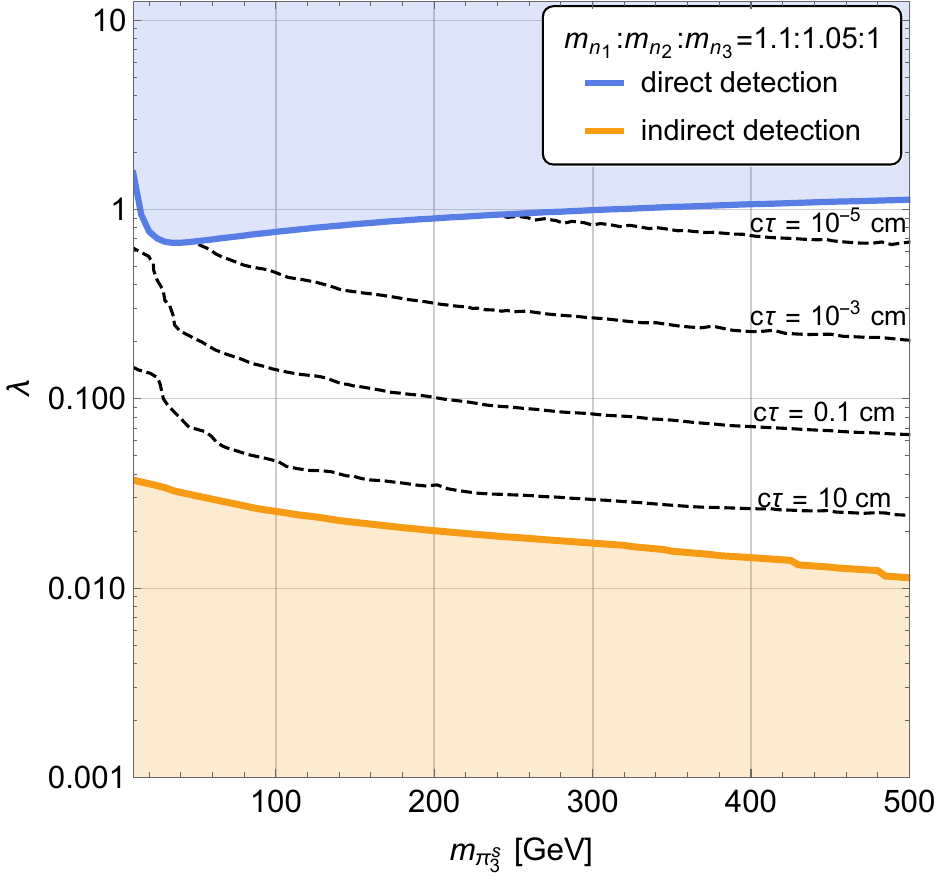}
    \caption{}
    \label{sfig:stranged}
  \end{subfigure}
\caption{Constraints on the coupling $\lambda^S_{D^c_2}$ and the mass of the lightest stable dark pion $m_{\pi_3^s}$. In each point the relic abundance matches the experimental value and $m_{X}=1$ TeV. The left panels show the contours of $m_{\pi^s_3}/f$, while the right ones have contours of the decay length of the lightest unstable dark pion $\pi_i^u$. In the upper panels we assumed a ratio of the dark quark masses of $n_1$:$n_2$:$n_3$=1.01:1.005:1, while in the lower panels we assumed $n_1$:$n_2$:$n_3$=1.1:1.05:1. The blue shaded area is excluded by direct detection experiments, while the orange region is excluded by indirect dark matter searches.}\label{fig:strange}
\end{figure}

\begin{figure}[t!]
  \centering
  \begin{subfigure}{0.48\textwidth}
    \centering
    \includegraphics[width=\textwidth, viewport = 0 0 450 425]{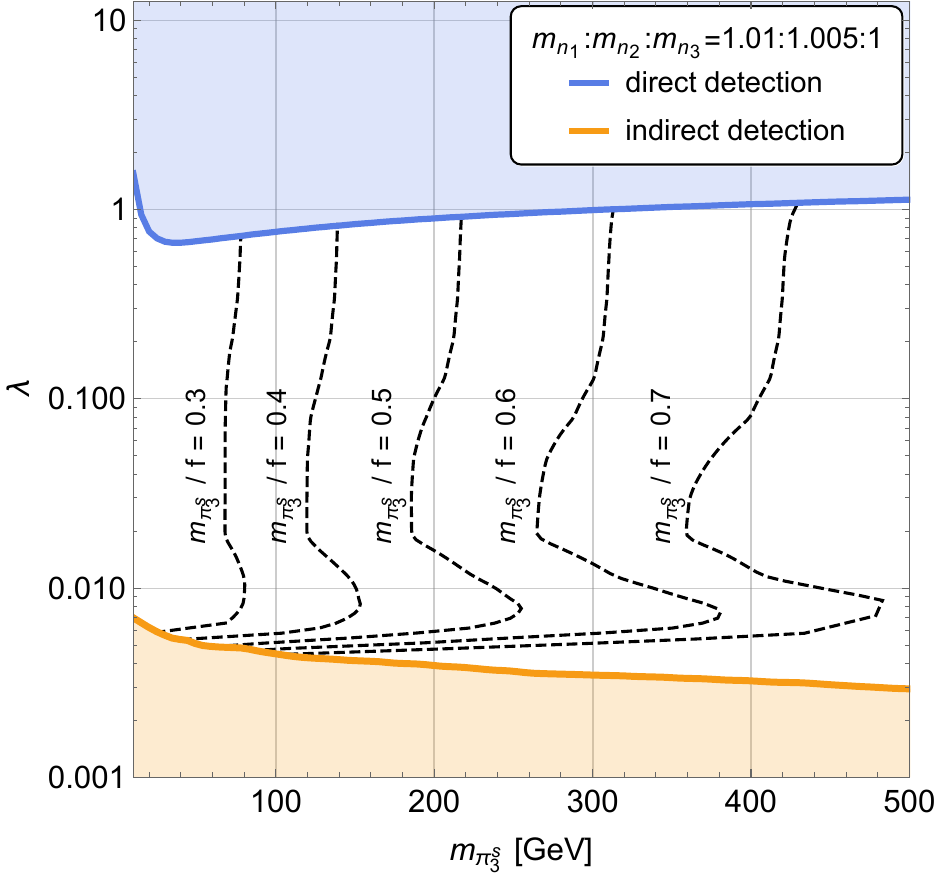}
    \caption{}
    \label{sfig:bottoma}
  \end{subfigure}
  ~
    \begin{subfigure}{0.48\textwidth}
    \centering
    \includegraphics[width=\textwidth, viewport = 0 0 450 425]{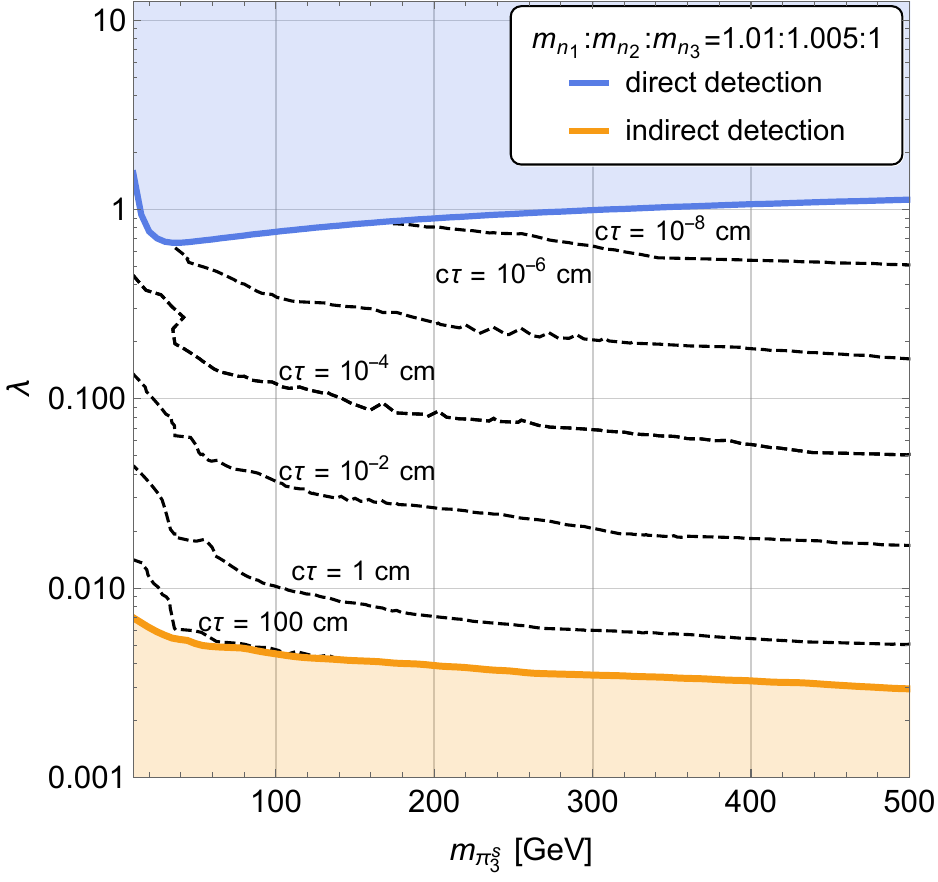}
    \caption{}
    \label{sfig:bottomb}
  \end{subfigure}
  ~
  \begin{subfigure}{0.48\textwidth}
    \centering
    \includegraphics[width=\textwidth, viewport = 0 0 450 425]{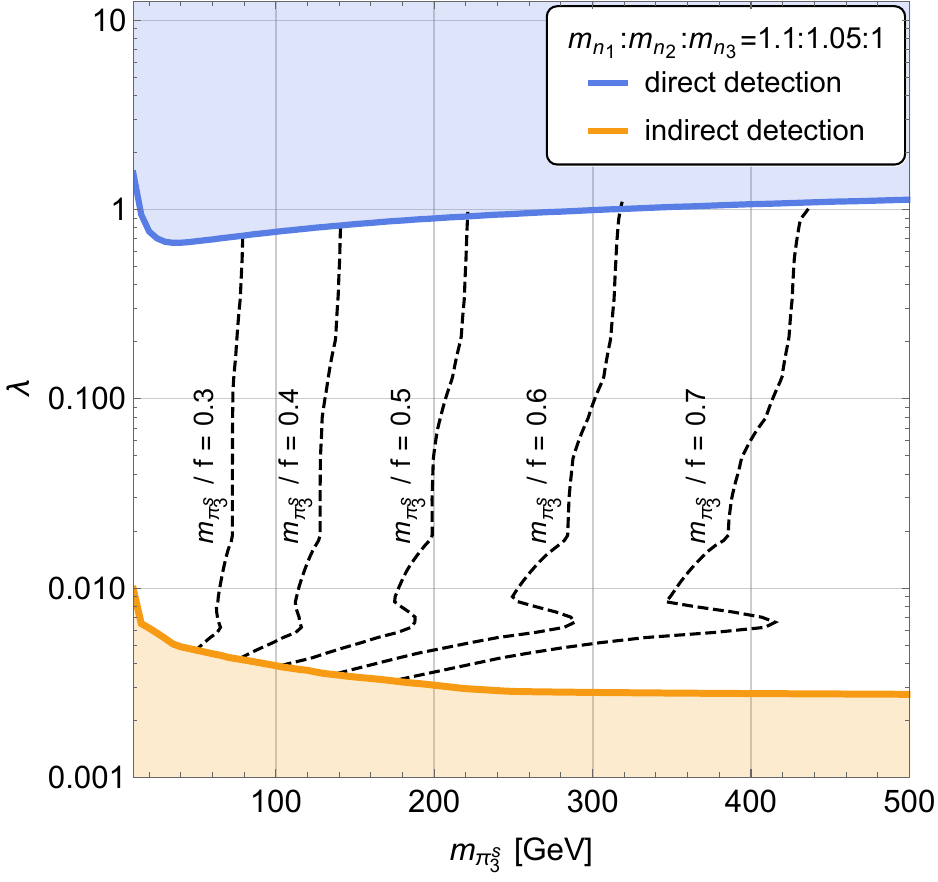}
    \caption{}
    \label{sfig:bottomc}
  \end{subfigure}
  ~
    \begin{subfigure}{0.48\textwidth}
    \centering
    \includegraphics[width=\textwidth, viewport = 0 0 450 425]{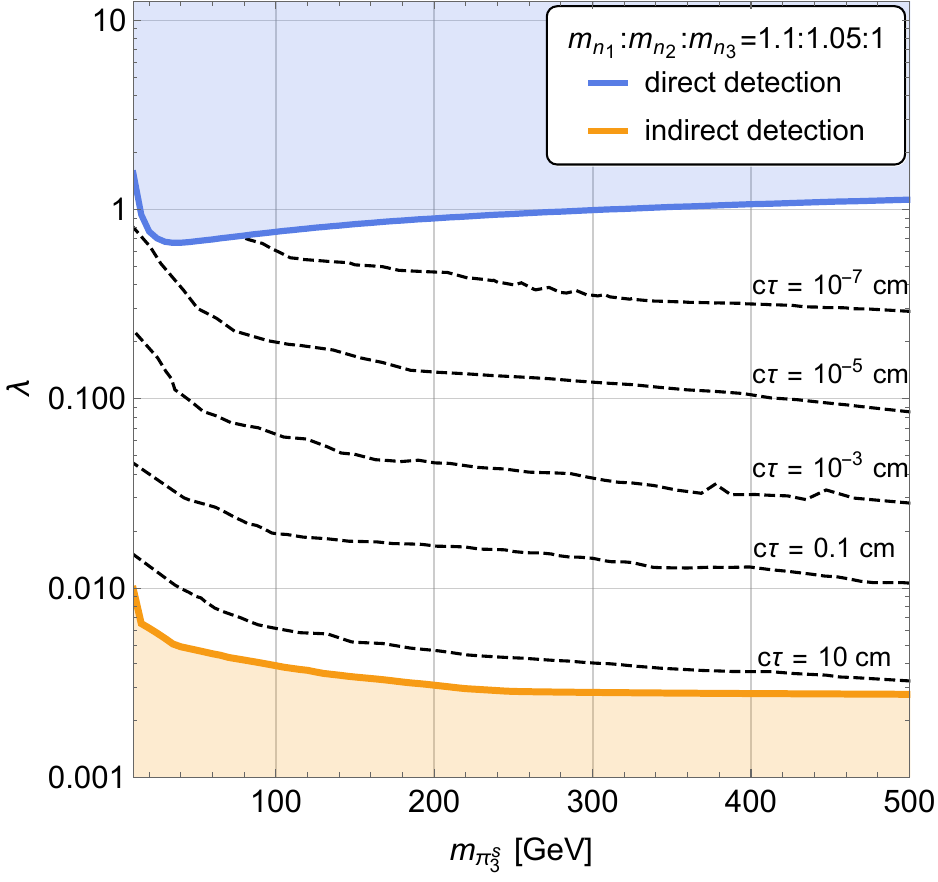}
    \caption{}
    \label{sfig:bottomd}
  \end{subfigure}
\caption{Constraints on the coupling $\lambda^S_{D^c_3}$ and the mass of the lightest stable dark pion $m_{\pi_3^s}$. In each point the relic abundance matches the experimental value and $m_{X}=1$ TeV. The left panels show the contours of $m_{\pi^s_3}/f$, while the right ones have contours of the decay length of the lightest unstable dark pion $\pi_i^u$. In the upper panels we assumed a ratio of the dark quark masses of $n_1$:$n_2$:$n_3$=1.01:1.005:1, while in the lower panels we assumed $n_1$:$n_2$:$n_3$=1.1:1.05:1. The blue shaded area is excluded by direct detection experiments, while the orange region is excluded by indirect dark matter searches.}\label{fig:bottom}
\end{figure}

Constraints are presented as a function of $\lambda^S_{D^c_i}$ and the mass of the lightest stable dark pion $m_{\pi^s_3}$ for coupling with down (Fig. \ref{fig:down}), strange (Fig.~\ref{fig:strange}) or bottom quarks (Fig. \ref{fig:bottom}). For each benchmark model we assume two different ratios of the dark quark masses: 1.01:1.005:1 or 1.1:1.05:1. Such ratios are not technically unnatural. The parameter $B_0$ of Eq.~\ref{eq:Chi} is taken equal to the mass of $n_3$. In all cases, the mediator mass is set to 1 TeV, which is still allowed by LHC constraints \cite{Beauchesne:2017yhh}. Since the mediator is integrated out, it is a trivial task to rescale most results for mediators of other masses. The pion decay constant $f$ is adjusted at each point to reproduce the correct dark matter relic density. 

Direct detection provides an upper limit on $\lambda^S_{D^c_i}$ for obvious reasons. As $\lambda^S_{D^c_i}$ decreases, codecaying dark matter eventually becomes the dominant mechanism, which leads to more dark matter for a fixed $f$. The pion decay constant must then be reduced to maintain the dark pions in equilibrium with each other for a longer time and thus reproduce the correct dark matter density. Eventually, this comes into conflict with indirect detection limits, which puts a lower bound on $\lambda^S_{D^c_i}$. Do note that there would also be a lower bound on the mass of the dark pion coming from its decay to SM fermions becoming forbidden. A proper treatment of this would however require considering hadronic degrees of freedom in some cases, which is beyond the scope of this article.

As can be seen from Fig.~\ref{fig:down}, the parameter space for dark pions decaying to down quarks is mostly ruled out, but there is still some viable parameter space. Mediators communicating with bottom quarks still have plenty of valid parameter space (Fig.~\ref{fig:bottom}), while mediators communicating with strange quarks represent a middle ground (Fig.~\ref{fig:strange}). The fact that there is more valid parameter space for heavier quarks is explained by two reasons. First, the upper bound from direct detection is obviously weaker for higher generations. Second, helicity suppression reduces the decay width to lighter fermions, which means that the system enters the codecaying regime for larger couplings and in turn means that bounds from indirect detection are stronger.

Two sets of contour lines are provided for Figs.~\ref{fig:strange} and \ref{fig:bottom}. In the left column, contour lines of $m_{\pi^s_3}/f$ are presented. As can be seen, this ratio increases as $m_{\pi^s_3}$ increases. This is simply because, as $m_{\pi^s_3}$ increases, the pions must remain in chemical equilibrium with each other until a larger $x$ to reproduce the correct dark matter relic density. Eventually, the approximation of keeping only the pseudo-Goldstone mesons would break down and additional dark hadrons would have to be included.\footnote{See Ref.~\cite{Berlin:2018tvf,Choi:2018iit} for the consequences of including vector mesons in the SIMP case.} For dark pions of a few hundred GeV or less, our formalism is still a very good approximation though. It also effectively places an upper limit on the mass of the dark pions, since we expect our computation to break at the latest for $m_{\pi^s_3}/f \gtrsim4 \pi$. Note that the region where these contour lines are vertical corresponds roughly to the coupling-independent regime. Fig.~\ref{sfig:strangea} and~\ref{sfig:strangec} and Fig.~\ref{sfig:bottoma} and~\ref{sfig:bottomc} show also how the system enters at larger couplings in the codecaying dark matter regime for lighter quarks.

It is worth noting that the results shown in Fig.~\ref{sfig:bottoma} and~\ref{sfig:bottomc} contain one additional feature. After the coupling-independent regime but before entering in the codecaying regime, the ratio $m_{\pi^s_3}/f$ decreases while $\lambda$ decreases. This is due to the fact that for $\lambda$ small, the energy exchange becomes less efficient and the dark pions become warmer with respect to the Standard Model sector because of the $3\to2$ processes. As a consequence, the relic abundance is below the experimental value and $f$ needs to increase in order to compensate.

The contour lines in the right column of Fig.~\ref{fig:strange} and Fig.~\ref{fig:bottom} represent the lifetime times the speed of light $c\tau$ of the lightest $\pi^u_i$. As can be seen, $c\tau$ can vary over such a range as to allow anything from dark pions decaying promptly to them escaping the detector. Dark mesons decaying promptly lead to semivisible jets, decays inside the detector lead to either displaced vertices or emerging jets depending on the multiplicity and decays outside the detector simply lead to missing transverse energy.

It is also interesting to note that the unexcluded range of $\lambda^S_{D^c_i}$ increases with $m_{\pi^s_3}$, but not as fast as one might expect. This has to do with the fact that there are competing effects that tend to increase and decrease the bounds from indirect detection. Schematically, pion scattering cross sections go as $m_{\pi^s_3}^2/f^4$. These cross sections decrease as $m_{\pi^s_3}$ increases, but not very fast as $m_{\pi^s_3}/f$ actually increases. At the same time, limits on the cross section become less important as $m_{\pi^s_3}$ increases. The net result is an indirect detection bound that only decreases slightly when $m_{\pi^s_3}$ increases, as can be seen in Figs.~\ref{fig:down}, \ref{fig:strange} and \ref{fig:bottom}.

Changing the ratio between the masses of the dark quarks can affect the bounds in potentially different ways. On one hand, increasing the splitting between the dark quarks increases the cross section of a stable pion and its antiparticle to unstable pions, which by itself would increase the bounds. On the other hand, it can affect the relative abundance of each type of stable dark pions, modifying the effective cross section relevant for indirect detection. The combination of these two effects is non-trivial and depends on a case by case basis. Larger splittings are not expected to change the results qualitatively.

\subsection{Multiple scalar mediators interacting only with down-type quarks}\label{sSec:ModelII}
Though instructive, the model of the previous section is not very realistic, as it is difficult to justify the dark pions interacting with one generation of quarks but not whatsoever with the others. It also does not exhibit certain features that would be present if dark pions were allowed to interact with multiple generations. In light of this, we present a model that involves all three generations of down-type quarks.

We assume that there are now three generations of mediators, which we now label $X^S_{D^c_k}$. The relevant Lagrangian term becomes:
\begin{equation}\label{eq:OpModelII}
  \lambda^S_{D^c_{ijk}} (X^S_{D^c_k})^\dagger \bar{n}_i P_R D^c_j +\text{h.c.} 
\end{equation}
We assume that each $X^S_{D^c_k}$ communicates with only a single generation of down-type quarks, that this generation is different for every mediator and that they all communicate only with $n_1$. In other words, we assume $\lambda^S_{D^c_{ijk}}\propto \delta_{i1}\delta_{jk}$. This is similar to the situation in supersymmetry, where down-type squarks are assumed to interact mainly with their corresponding generation, even though this need not be the case in general. Having the mediators couple only to $n_1$ is technically natural.

\begin{figure}[t!]
  \centering
  \begin{subfigure}{0.48\textwidth}
    \centering
    \includegraphics[width=\textwidth, viewport = 0 0 450 425]{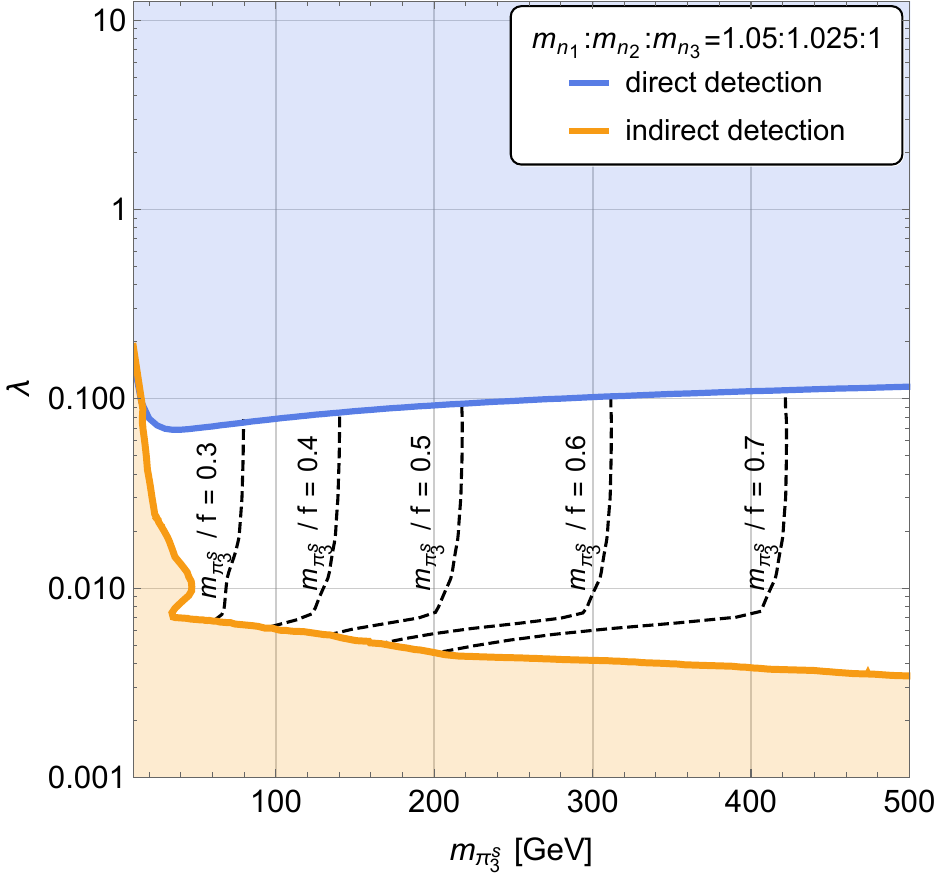}
    \caption{}
    \label{sfig:Alldowna}
  \end{subfigure}
  ~
    \begin{subfigure}{0.48\textwidth}
    \centering
    \includegraphics[width=\textwidth, viewport = 0 0 450 425]{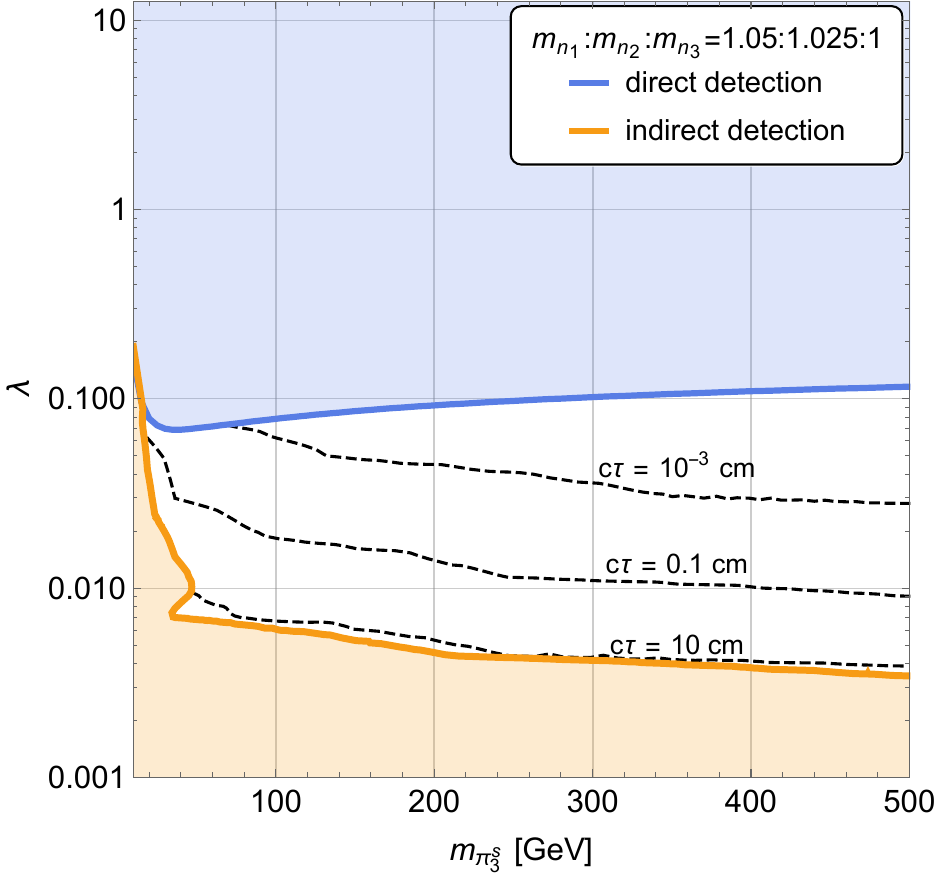}
    \caption{}
    \label{sfig:Alldownb}
  \end{subfigure}
\caption{Constraints on the coupling $\lambda$ and the mass of the lightest stable dark pion $m_{\pi_3^s}$. In each point the relic abundance matches the experimental value and $m_{X}=1$ TeV. The left panel shows the contours of $m_{\pi^s_3}/f$, while the right one has contours of the decay length of the lightest unstable dark pion $\pi_i^u$. In both panels we assumed a ratio of the dark quark masses of $n_1$:$n_2$:$n_3$=1.05:1.025:1. The blue shaded area is excluded by direct detection experiments, while the orange region is excluded by indirect dark matter searches.}\label{fig:Alldown}
\end{figure}

Constraints are presented in Fig.~\ref{fig:Alldown} for $\lambda^S_{D^c_{ijk}} = \lambda\delta_{i1}\delta_{jk}$, mediators of degenerate masses and ratio of the dark quark masses of 1.05:1.025:1. As can be seen, the constraints contain features of the model of Sec.~\ref{sSec:ModelI} for both light and heavy generations. The bound from direct detection is strong, as the dark pions couple to the down quarks directly. Conversely, the bounds from indirect detection is at relatively small $\lambda$. This is because the dark pions decay mostly to bottom quarks due to helicity suppression and this allows for a large region of parameter space to fall under the coupling-independent regime.

Of course, the assumption of $\lambda^S_{D^c_{ijk}} = \lambda\delta_{i1}\delta_{jk}$ can be relaxed and the consequences of this are trivial. Reducing the coupling to the first generation simply means that tensions with direct detection are reduced. Reducing the coupling to the third generation simply means that the bound on indirect detection is increased. The impact on direct and indirect detection of the other $\lambda^S_{D^c_{ijk}}$ is generally subdominant. Helicity suppression means that dark jets at colliders will contain mostly heavy flavor hadrons.

\subsection{$Z'$ model}\label{sSec:ModelIII}
A Hidden sector dark quark can interact with the Standard Model sector via a TeV-scale $Z'$. Since they are a valid alternative to the bifundamental scalar studied in the previous sections, we consider the extension of the Standard Model by an extra broken $U(1)$ gauge symmetry. We take the gauge coupling of this new group to be $\hat{g}$ and assume all SM fermion fields to have a $U(1)$ charge equal to their weak hypercharge to make the theory anomaly free \cite{Appelquist:2002mw}. We assume $n_1$ to have a charge of 1, but take $n_2$ and $n_3$ to be neutral. The ratios of the dark quark masses are taken as 1.05:1.025:1. 

\begin{figure}[t!]
      \centering
  \begin{subfigure}{0.48\textwidth}
    \centering
    \includegraphics[width=\textwidth, viewport = 0 0 450 425]{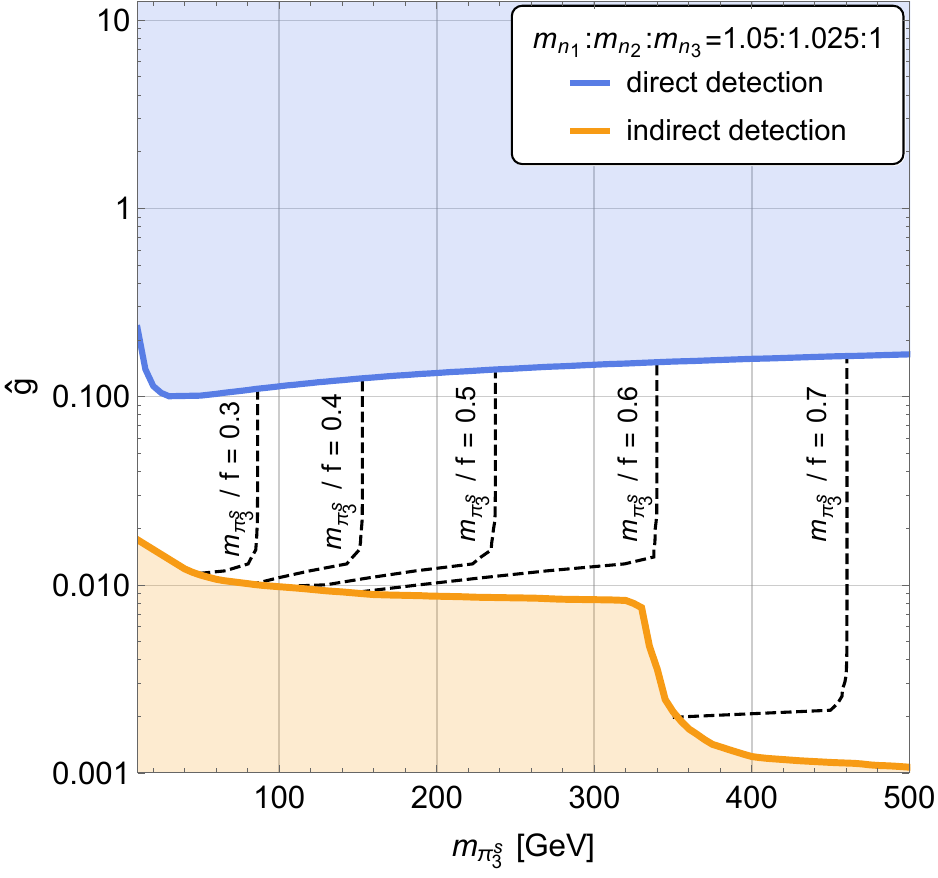}
    \caption{}
     \label{sfig:Zprimea}   
  \end{subfigure}
  ~
  \begin{subfigure}{0.48\textwidth}
    \centering
    \includegraphics[width=\textwidth, viewport = 0 0 450 425]{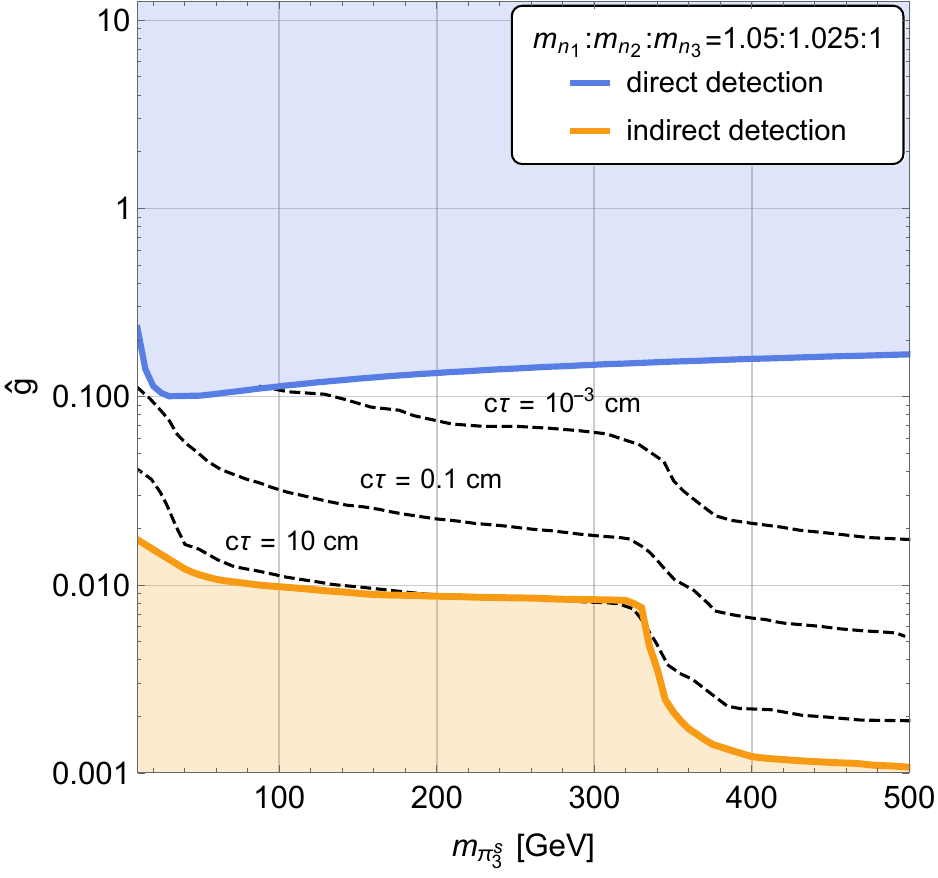}
    \caption{}
     \label{sfig:Zprimeb} 
  \end{subfigure}
\caption{Constraints on the coupling $\hat{g}$ and the mass of the lightest stable dark pion $m_{\pi_3^s}$. In each point the relic abundance matches the experimental value and $m_{Z'}=3$ TeV. The left panel shows the contours of $m_{\pi^s_3}/f$, while the right one has contours of the decay length of the lightest unstable dark pion $\pi_i^u$. In both panels we assumed a ratio of the dark quark masses of $n_1$:$n_2$:$n_3$=1.05:1.025:1. The blue shaded area is excluded by direct detection experiments, while the orange region is excluded by indirect dark matter searches.}\label{fig:Zprime}
\end{figure}

Results are shown in Fig.~\ref{fig:Zprime} as a function of $m_{\pi^s_3}$ and $\hat{g}$: the left panel shows also contours of $m_{\pi_{1}^s}/f$, while the right one shows the decay length $c\,\tau$ of $\pi_i^u$. In the plot we chose a $Z'$ mass of $m_{Z'} = 3$ TeV in order to avoid current collider constraints \cite{Ekstedt:2016wyi}. The direct detection constraints are strong, due to the fact that we have a direct coupling between the $Z'$ and the up and down quark, ruling out couplings of order $\mathcal{O}(0.1)$ for all the considered dark pion masses. In contrast, the indirect detection constraints are very strong in the parameter space where the dark pion decay to two top quarks is  kinematically forbidden, excluding $\hat{g} \lesssim 10^{-2}$. However, when the decay to top quarks is allowed, the coupling-independent region extend to lower couplings and therefore the indirect detection constraints become weaker, excluding only the parameter space below $\hat{g} \lesssim 10^{-3}$.

\section{Additional comments}\label{Sec:Comments}
In this section, we present a few additional points worth mentioning.

\begin{figure}[t!]
      \centering
  \begin{subfigure}{0.48\textwidth}
    \centering
    \includegraphics[width=\textwidth, viewport = 0 0 450 425]{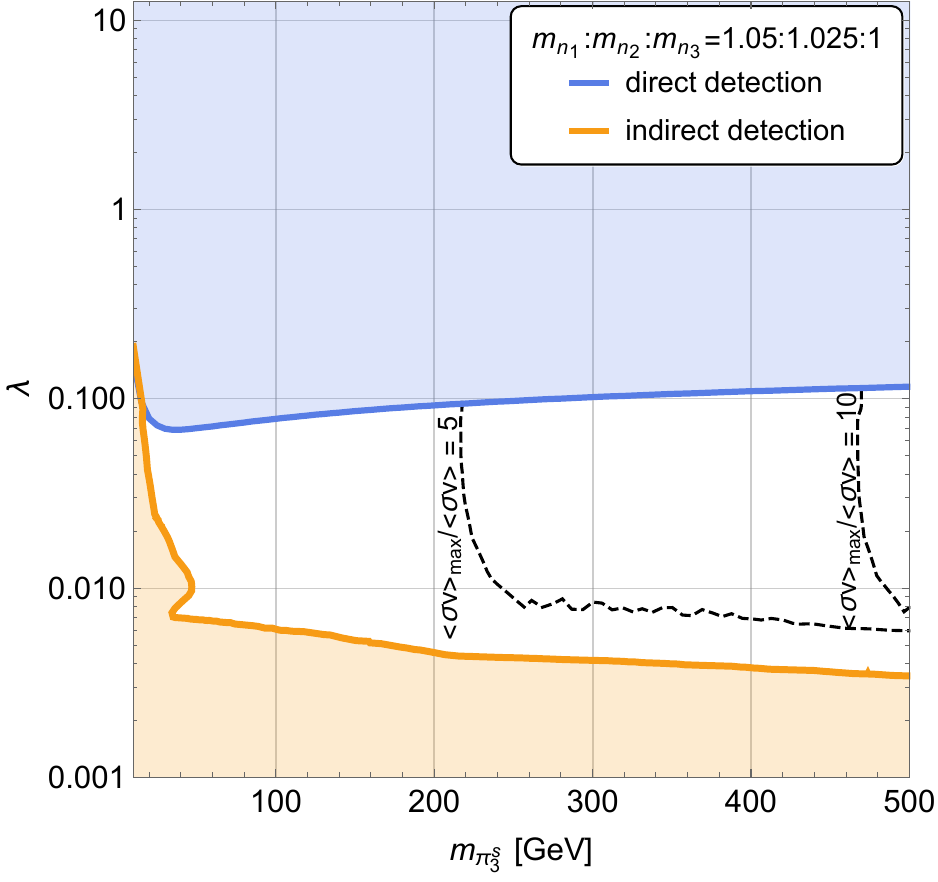}
    \caption{}
     \label{sfig:sigmava}   
  \end{subfigure}
  ~
  \begin{subfigure}{0.48\textwidth}
    \centering
    \includegraphics[width=\textwidth, viewport = 0 0 450 425]{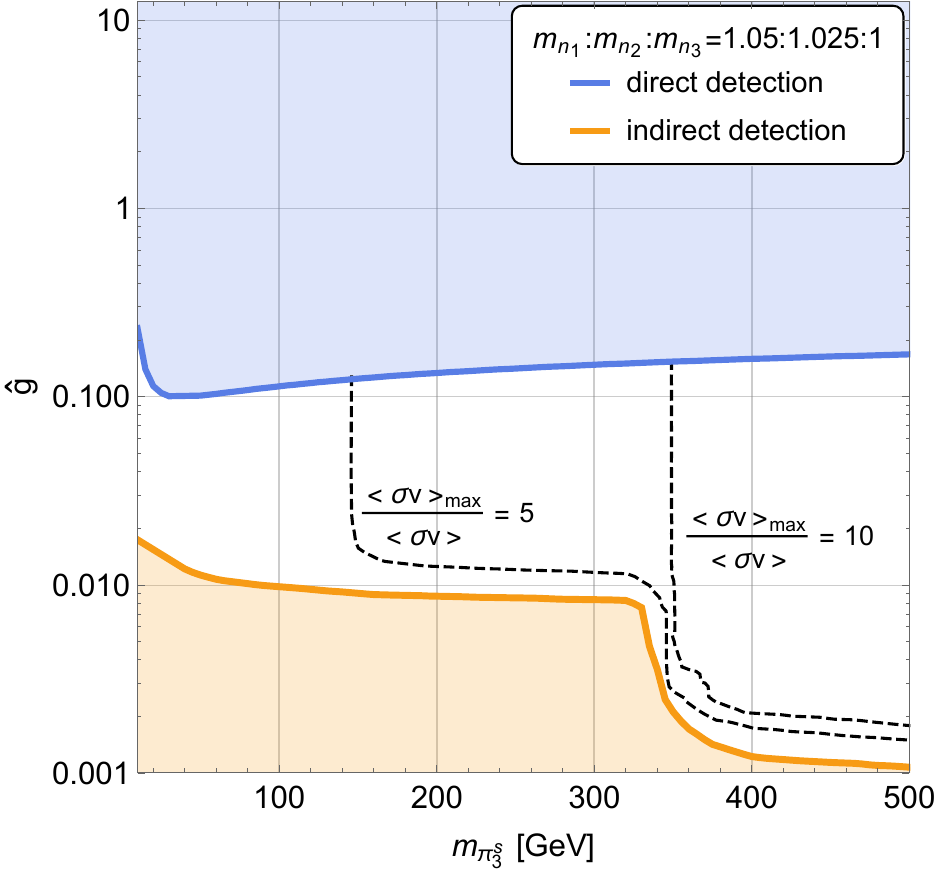}
    \caption{}
     \label{sfig:sigmavb} 
  \end{subfigure}
\caption{Constraints on the coupling $\lambda$ (a) and $\hat{g}$ (b) vs the mass of the lightest stable dark pion $m_{\pi_3^s}$ with the same assumptions of the previous plots.The blue shaded area is excluded by direct detection experiments, while the orange region is excluded by indirect dark matter searches. The plot shows also the contours of $\langle \sigma v \rangle_{\mathrm{max}} / \langle \sigma v \rangle$.}\label{fig:sigmav}
\end{figure}

First, the main reason for having so much parameter space currently unexcluded is because a very large portion of it falls under the coupling-independent regime and that this regime is generally just below the bounds from indirect detection. Conversely, if indirect detection cross section limits were to increase significantly, they would exclude a very large part of the parameter space. This is illustrated in Fig.~\ref{fig:sigmav}, which shows by how much the limit on the indirect detection cross section would have to increase to exclude a given point of parameter space for the model of Sec.~\ref{sSec:ModelII} and for the $Z'$ model of Sec.~\ref{sSec:ModelIII}. As one can see, very large parts of the parameter space lay within an order of magnitude. In particular an increase in the sensitivity of indirect detection searches by a factor 10 would be able to probe a large region of the  model of Sec.~\ref{sSec:ModelII} for dark pion masses below about $480$ GeV and the $Z'$ model studied in this work for dark pions lighter than about $360$ GeV.\footnote{On the other hand, an improvement of one order of magnitude in the sensitivity on the spin independent cross section would lead to an improvement of only a factor of about two in the coupling $\lambda$ ($\hat{g}$). This is because the spin independent cross section is independent of the value of $f$ at leading order.}  Hence, if dark matter really consists of the stable mesons of a dark sector that also contains unstable ones, it should be detectable soon by indirect detection. 

Second, it is worth mentioning that the presence of the coupling-independent regime allows the generation of the correct dark matter relic abundance for masses of the Hidden Valley mediators varying by orders of magnitude. If these can be produced on-shell at colliders, they can potentially have sizeable cross sections, as for the models of Sec~\ref{sSec:ModelI} and \ref{sSec:ModelII}, which means an abundant production of dark quarks and hence dark pions.\footnote{See Ref.~\cite{Beauchesne:2017yhh} for some example cross sections.} If the mediators cannot be produced on-shell, the production of dark quarks will be suppressed by some power of some potentially small $\lambda$, which means a smaller cross section for the production of dark pions. This would make probing the very small $\lambda$ region very difficult at colliders, though we leave the details for a future analysis.

\begin{figure}[t!]
      \centering
  \begin{subfigure}{0.48\textwidth}
    \centering
    \includegraphics[width=\textwidth, viewport = 0 0 450 425]{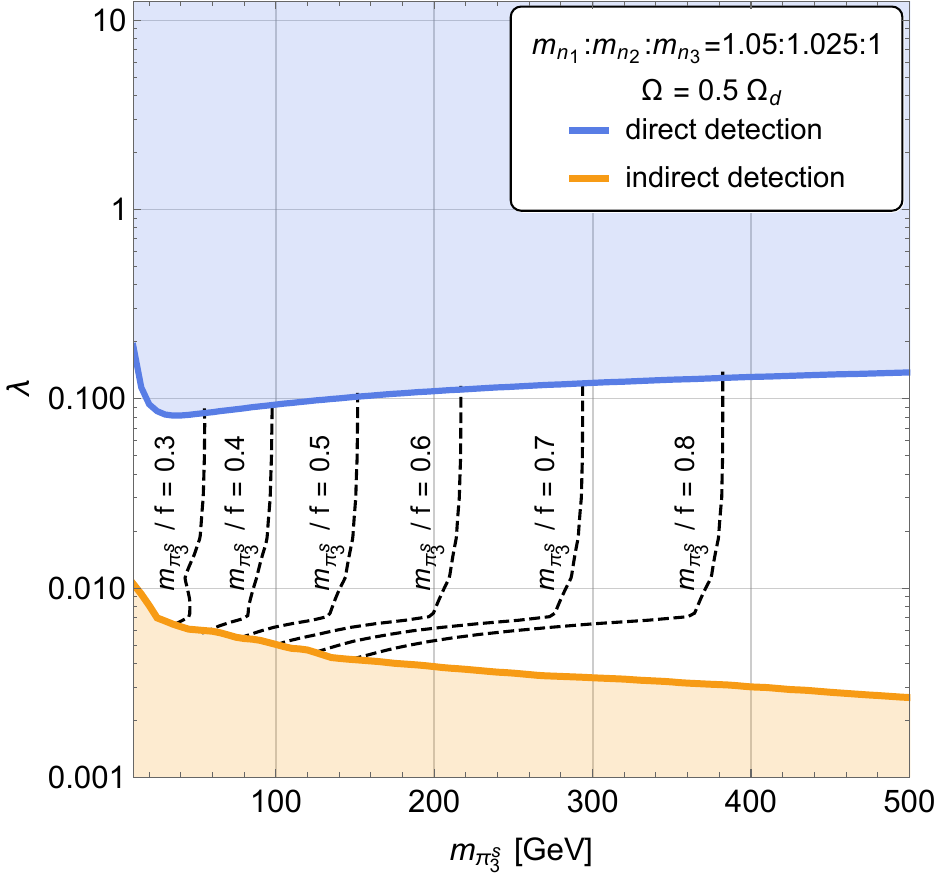}
    \caption{}
     \label{sfig:omegaa}   
  \end{subfigure}
  ~
  \begin{subfigure}{0.48\textwidth}
    \centering
    \includegraphics[width=\textwidth, viewport = 0 0 450 425]{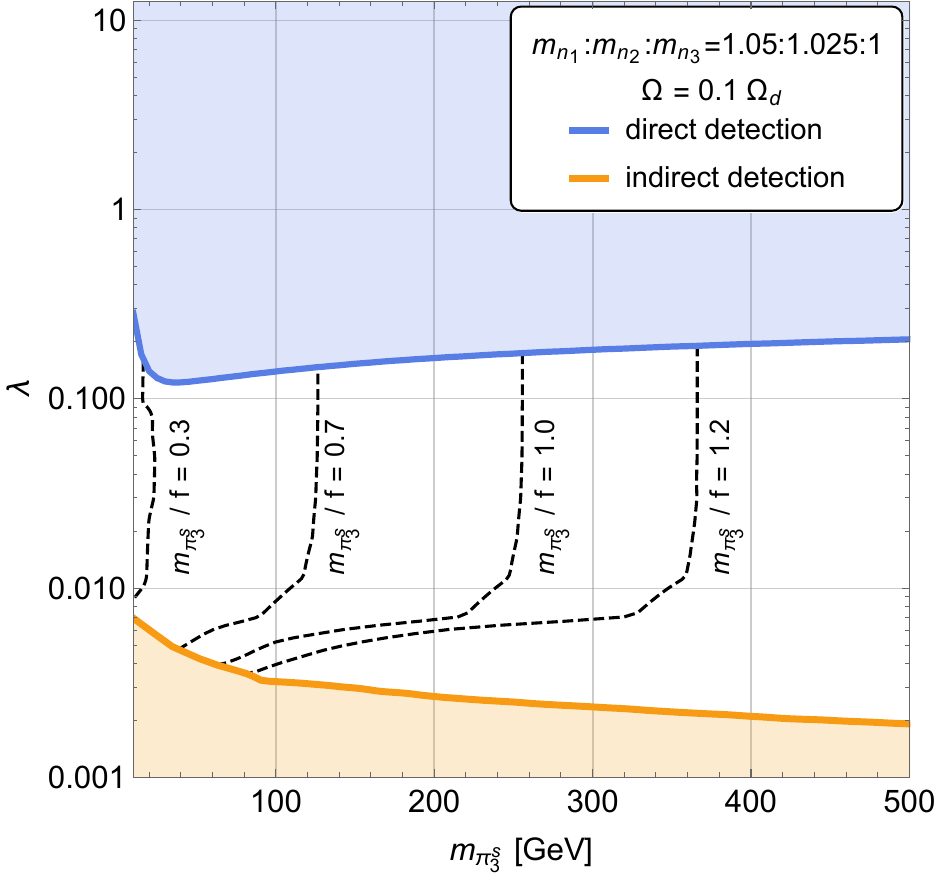}
    \caption{}
     \label{sfig:omegab} 
  \end{subfigure}
\caption{Constraints on the coupling $\lambda$ and the mass of the lightest stable dark pion $m_{\pi_3^s}$. The left (right) panel shows the bounds for dark pions representing 50\% (10\%) of the total dark matter abundance. In both panels we assumed $m_{X}=1$ TeV, a ratio of the dark quark masses of $n_1$:$n_2$:$n_3$=1.05:1.025:1 and we show the contours of $m_{\pi^s_3}/f$. The blue shaded area is excluded by direct detection experiments, while the orange region is excluded by indirect dark matter searches.}\label{fig:omega}
\end{figure}

Third, it certainly would not be surprising if a new confining sector were to contain stable particles beyond the dark pions. These could for example be dark baryons in the presence of an asymmetry. As such, we present in Fig.~\ref{fig:omega} the allowed parameter space for the dark pions representing 50$\%$ (left) or 10$\%$ (right) of the total dark matter relic abundance. 
The model of Sec.~\ref{sSec:ModelII} was used. The bounds from direct detection are slightly less stringent, as the scattering amplitudes of dark pions with matter are independent of $f$ but the dark pion density is reduced. In particular, we compute the direct detection rates rescaling the local DM density according to the prescription $\rho = \rho_{\mathrm{loc}}\, \mathrm{min}\left(1,\frac{\Omega}{\Omega_d}\right)$, where $\Omega$ is the relic density of the dark pions. As a consequence, the bounds on the coupling varies by a factor $\mathrm{max}\left(1,\frac{\Omega_d}{\Omega}\right)^{1/4}$. On the other hand, the bounds from indirect detection are mostly similar. This is because the reduction in the dark pions abundance is compensated by the dark pions having to interact more with each other to decrease their abundance to sufficient levels.

\section{Conclusion}\label{Sec:Conclusion}
In this article, we studied Hidden Valley sectors whose spectra of light particles consist of both stable and unstable dark mesons. It was shown that these sectors can reproduce the correct dark matter relic abundance, while easily avoiding the bounds from direct and indirect detection. In most of the valid parameter space, the mechanism that determines the dark matter relic abundance is the annihilation of two stable dark pions to two unstable ones that decay quickly. This mechanism is responsible for a large region of parameter space where the dark matter relic abundance is essentially independent of the couplings between the dark pions and the Standard Model, which results in much of the parameter space being allowed by experimental constraints. Such Hidden Valley sectors can conversely lead to very exotic signatures at colliders, such as displaced vertices, emerging jets and semivisible jets. We also showed that much of the parameter space would be probeable by increasing the limits of indirect detection on dark matter annihilation cross section by roughly an order of magnitude. Finally, the confining sector may contain dark baryons in the presence of an asymmetry and therefore the dark pions could be only a fraction of the total dark matter relic abundance. As a consequence, we showed how the bounds from direct and indirect dark matter detection experiments are modified with this assumption.

\acknowledgments
This work was supported by Fundação de Amparo à Pesquisa (FAPESP), under contracts 16/17040-3, 16/17041-0 and 15/25884-4, and Conselho Nacional de Ci\^{e}ncia e Tecnologia (CNPq). HB is grateful to the Azrieli Foundation for the award of an Azrieli Fellowship.

\appendix

\section{Calculations of the dark matter relic density}\label{ap:Technicalities}
In this appendix, we summarize the different results that were used to calculate the dark matter relic densities. Most of them are standard in the literature and are provided as a reference for the reader. 
\subsection{Lagrangian}\label{sap:Lagrangian}
Interactions between the dark quarks and the Standard Model are encoded in the effective Lagrangian
\begin{equation}\label{eq:LagrangianDarkQuarks}
  \mathcal{L}^n = i\bar{n}\slashed{D}n -\bar{n}\gamma_\mu\left(l^\mu P_L + r^\mu P_R\right) n - \bar{n}(s - i p \gamma^5)n,
\end{equation}
where $n$ represents the array of the $n_i$'s. The fields $l$, $r$, $s$ and $p$ are linear combinations of bilinears of Standard Model fermions and are obtained by integrating out the mediator. They have in principle two indices associated to the dark quark generation, but in all cases we consider only a single element of the diagonal is non-zero. Note that the mass of the dark quarks is included in our definition of $s$. In terms of dark pions degrees of freedom, the interactions with the Standard Model can be described by the Lagrangian
\begin{equation}\label{eq:LagrangianII}
 \mathcal{L}^\Pi  = \frac{f^2}{4}\text{Tr}\left[D_\mu U (D^\mu U)^\dagger + \chi U^\dagger + U \chi^\dagger \right],
\end{equation}
where $f$ is the pion decay constant,
\begin{equation}\label{eq:defU}
  U = \text{exp}\left[i\frac{\sqrt{2}}{f}\Pi\right],
\end{equation}
the covariant derivatives are defined as
\begin{equation}\label{eq:CovariantDeriviative}
 D_\mu U = \partial_\mu U + i l_\mu U - i U r_\mu
\end{equation}
and
\begin{equation}\label{eq:Chi}
  \chi = 2 B_0(s + i p),
\end{equation}
where $B_0$ is a constant that relates the mass of the dark quarks to the mass of the dark pions.

\subsection{$3\to 2$ processes}\label{sap:3to2}
In addition to the Lagrangian of the previous section, the full Lagrangian describing dark pion interactions includes the Wess-Zumino-Witten term, given at leading order by \cite{Hochberg:2014kqa}\footnote{Note that the definition of $f$ differs by a factor of 2 with respect to Ref.~\cite{Hochberg:2014kqa} and $\Pi$ by a factor of $\sqrt{2}$.}
\begin{equation}\label{eq:WZWI}
  \mathcal{L}_{\text{WZW}} = \frac{N_c}{15\sqrt{2}\pi^2 f^5}\epsilon^{\mu\nu\alpha\beta}\text{Tr}\left[\Pi \partial_\mu\Pi \partial_\nu\Pi \partial_\alpha\Pi \partial_\beta\Pi  \right],
\end{equation}
where $N_c =3$ is the number of dark colors. This term can be rearranged as \cite{Hochberg:2014kqa}
\begin{equation}\label{eq:WZWII}
  \mathcal{L}_{\text{WZW}} = \frac{N_c}{240\pi^2 f^5}\epsilon^{\mu\nu\alpha\beta} \sum_{a < b < c < d < e} T_{abcde} \pi_a \partial_\mu\pi_b \partial_\nu\pi_c \partial_\alpha\pi_d \partial_\beta\pi_e,
\end{equation}
where $\pi_a$ represents the set of all pions and $T_{abcde}$ is a set of group theoretic constants. This term is responsible for the $3\to 2$ scattering processes, whose thermally-averaged cross section is given for degenerate pions at leading order by \cite{Hochberg:2014kqa}
\begin{equation}\label{eq:3to2ThermalCrossSection}
  \langle \sigma v^2\rangle^{abc\to de}_\pi = \frac{N_c^2m_\pi^3 T'^2}{2^{15}3 \sqrt{5}\pi^5 f^{10}}T_{\{abcde\}}^2,
\end{equation}
where $T'$ is the temperature of the dark pions sector and $\{abcde\}$ means the normal-ordering of $a$, $b$, $c$, $d$ and $e$. Note that the basis of Eq.~\ref{eq:PionMatrix} makes it trivial to see which scattering processes should be zero because of conservation of $U(1)_i$ charges. As this is a $3\to 2$ process, this result is still valid to good approximation for non-degenerate pions as long as the difference between the pion masses are small with respect to their average mass. Since we only consider small splittings, we therefore use this result directly even when the pions are not degenerate.

\subsection{Energy exchange}\label{sap:EnergyExchange}
In addition to changing the number of dark matter particles, interactions between the Standard Model and the Hidden Valley sectors are also responsible for exchanging energy and thus maintaining kinetic equilibrium. If the coefficient of the effective operator controlling this exchange is very small, it is possible that kinetic exchange will become inefficient before the amount of dark matter has settled to its final value. As such, it is possible that the dark pions temperature $T'$ differs from that of the SM sector $T$, which means that $T'$ must be tracked. This is done by adding another differential equation to the Boltzmann equations of Sec.~\ref{sap:BoltzmannEquations}. This differential equation will contain a term coming from scattering which we discuss here.

Assume a scattering process between a dark pion $\pi_i$ and a SM fermion $f$. The differential equation will contain the term \cite{Kuflik:2017iqs, Bringmann:2006mu}
\begin{equation}\label{eq:EnergyExchangeI}
  \begin{aligned}
    n_{\pi_i} n_f \langle \sigma v \delta E \rangle^i &= - \int d\Pi_{\pi_i} d\Pi_f d\Pi_{\pi_i'} d\Pi_{f'}(2\pi)^4 \delta^4\left(p_{\pi_i'} + p_{f'} - p_{\pi_i} - p_f\right)\\
                                                      &  \hspace{1.2cm} \times \frac{p_{\pi_i}^2}{2m_{\pi_i}}\left(f_{\pi_i}f_f - f_{\pi_i'}f_{f'}\right)\overline{\left|M_{\pi_i f \to \pi_i' f'}\right|}^2\\
                                                      &= - \int d\Pi_{\pi_i} d\Pi_f d\Pi_{\pi_i'} d\Pi_{f'}(2\pi)^4 \delta^4\left(p_{\pi_i'} + p_{f'} - p_{\pi_i} - p_f\right)\frac{p_{\pi_i}^2}{2m_{\pi_i}}C(T),
  \end{aligned}
\end{equation}
where primes indicate a given quantity after the collision, the $f_i$ are the phase-space occupancies and $C(T)$ is the collision term. Note that the occupancy number of SM particles are a function of $T$ and that of dark pions a function of $T'$. The collision term can be expressed as \cite{Kuflik:2017iqs, Bringmann:2006mu}
\begin{equation}\label{eq:EnergyExchangeII}
  C(T) = \frac{1}{768\pi^3 T m_{\pi_i}^2}\int_{m_f}^{\infty} d\omega f_f(\omega)\int_{-4k_{CM}^2}^0 dt \overline{\left|M_{\pi_i f \to \pi_i' f'}\right|}^2\left(m_{pi_i} T\Delta_{\mathbf p} + \mathbf{p}\cdot \nabla_{\mathbf p} +3 \right)f_{\pi_i}(\mathbf{p}),
\end{equation}
where $\mathbf{p}$ is the 3-momentum of the incoming dark pion and
\begin{equation}\label{eq:EnergyExchangeIII}
  k_{CM}^2 = \frac{\left(s - (m_{\pi_i} - m_f)^2\right)\left(s - (m_{\pi_i} + m_f)^2\right)}{4s}
\end{equation}
and
\begin{equation}\label{eq:EnergyExchangeIV}
  s = m_{\pi_i}^2 + 2m_{\pi_i} \omega + m_f^2.
\end{equation}

\subsection{Boltzmann equation and approximations}\label{sap:BoltzmannEquations}
The Boltzmann equation for the number density of dark pions $\pi_i$ is
\begin{equation}\label{eq:BoltzmanI}
  \begin{aligned}
    \frac{d Y_i}{dx} &= -\frac{c}{x^2}\left(  \sum_j \langle\sigma v\rangle_{\text{SM}}^{ij}\left(Y_i Y_j - Y_i^{\text{eq}} Y_j^{\text{eq}}\right) + \sum_j \langle\sigma v\rangle^j_{\text{conv}} \left(Y_i - \frac{Y_i^{\text{eq}}}{Y_j^{\text{eq}}} Y_j\right)Y_f^{\text{eq}} \right.\\
    & \hspace{1.5cm} + \frac{\langle\Gamma_i\rangle}{s}\left(Y_i - Y_i^{\text{eq}} \right) \\
    & \hspace{1.5cm}\left. + \sum_{j,m,n} \langle\sigma v\rangle_{\pi}^{ij \to mn}\left(Y_i Y_j - \frac{Y_i^{\text{eq}} Y_j^{\text{eq}}}{Y_m^{\text{eq}} Y_n^{\text{eq}}}Y_m Y_n  \right)  \right.\\
    & \hspace{1.5cm}\left. + \sum_{j,k,m,n} s\langle\sigma v^2\rangle_{\pi}^{ijk \to mn}\left(Y_i Y_j Y_k - \frac{Y_i^{\text{eq}} Y_j^{\text{eq}} Y_k^{\text{eq}}}{Y_m^{\text{eq}} Y_n^{\text{eq}}}Y_m Y_n \right)\right.\\
    & \hspace{1.5cm}\left. - \sum_{j,k,m,n} s\langle\sigma v^2\rangle_{\pi}^{jkm \to in}\left(Y_j Y_k Y_m - \frac{Y_j^{\text{eq}} Y_k^{\text{eq}} Y_m^{\text{eq}}}{Y_i^{\text{eq}} Y_n^{\text{eq}}}Y_i Y_n \right)\right).
  \end{aligned}
\end{equation}
A few explanations are in order. The parameter $Y_i$ is the number density per entropy density $n_i/s$ and $Y_i^{\text{eq}}$ its equilibrium value. The constant $c$ is given by
\begin{equation}\label{eq:BoltzmanII}
  c = \sqrt{\frac{\pi g_\ast(T)}{45G}}m_{\pi^s_1},
\end{equation}
where $g_\ast(T)$ is the number of relativistic degrees of freedom at temperature $T$. We neglect the difference between this quantity and the number of relativistic degrees of freedom that contribute to the entropy $g_{\ast s}(T)$.

The first two lines of Eq.~(\ref{eq:BoltzmanI}) contain terms that involve SM particles. The first term corresponds to annihilation of two dark pions to SM particles and $\langle\sigma v\rangle_{\text{SM}}^{ij}$ represents the thermally averaged cross section for that process. Stable pions have to annihilate with their antiparticle, while unstable ones can also annihilate between $\pi^u_1$ and $\pi^u_2$. The second term corresponds to conversion of pions by scattering of SM particles and $\langle\sigma v\rangle^j_{\text{conv}}$ corresponds to the associated cross section. This is only possible for $\pi^u_1$ going to $\pi^u_2$ or vice-versa. The third term corresponds to decay of the dark pions to SM particles. Obviously, this term is zero for stable pions. The quantity $\langle \Gamma_i\rangle$ refers to the decay width times the thermally averaged inverse gamma factor of $\pi_i$. If the dark pions can interact with multiple SM fermions, these must be summed over.

The third line refers to $2 \to 2$ pion scattering. The quantity $\langle\sigma v\rangle_{\pi}^{ij \to mn}$ is the relevant thermally averaged cross section. The fourth and fifth lines refer to $3 \to 2$ processes. The quantity $\langle\sigma v^2\rangle_{\pi}^{ijk \to mn}$ is the associated thermally averaged cross section.

The temperature of the dark sector $T'$ is tracked by using the evolution of its energy density. The differential equation governing the energy density $\rho_i$ of $\pi_i$ is
\begin{equation}\label{eq:BoltzmanIII}
  \frac{d\rho_i}{dt} + 3H\left(\rho_i + P_i\right) = - \langle \sigma v \delta E \rangle^i n_i n_f - m_i \Gamma_i\left(n_i - n_i^{\text{eq}}\right) + \ldots,
\end{equation}
where $P_i$ is the pressure and where the ellipsis refers to terms related to energy exchange between dark pions, which cancel when summed over all pions and which are irrelevant to the following discussion. An additional term related to annihilation of pions to SM particles could also be included. When the temperature of the sector start to diverge, this term will however become subdominant to the other terms in Eq.~(\ref{eq:BoltzmanIII}) and is therefore neglected. Possibly using one of the approximations that follow, Eq.~(\ref{eq:BoltzmanIII}) can be transformed in a differential equation for $T'$ as a function of $x$, in some cases with the help of Eq.~(\ref{eq:BoltzmanI})

Fully solving Eq.~(\ref{eq:BoltzmanIII}) for couplings varying by orders of magnitudes is a challenging tasks. As such, we make a series of approximations:
\begin{itemize}
  \item When the energy exchange rate between the SM and the Hidden sector divided by the dark pion mass is much higher that the Hubble constant, we approximate the temperatures of the two sectors as equal.
  \item When $2\to 2$ processes between dark pions occur at a rate much higher than the Hubble constant, we assume the dark pions share the same chemical potential.
  \item When $3\to 2$ processes between dark pions occur much faster than the Hubble constant, we assume the dark pions to be in chemical equilibrium.
\end{itemize}
Otherwise, Eqs.~(\ref{eq:BoltzmanI}) and Eqs.~(\ref{eq:BoltzmanIII}) are fully solved numerically.

\bibliography{biblio}
\bibliographystyle{JHEP}

\end{document}